\RequirePackage[final]{graphicx}
\documentclass[aps, superscriptaddress, prx, twocolumn]{revtex4-1}
\usepackage[utf8]{inputenc}
\usepackage{amsmath}
\usepackage{amssymb}
\usepackage{amsfonts}
\usepackage{mathtools}
\usepackage{fixme}
\usepackage{color}
\usepackage{braket}
\usepackage{dsfont}
\usepackage{hyperref}
\usepackage{url}

\usepackage{tikz}
\usetikzlibrary{calc}

\usetikzlibrary{positioning,arrows}
\usetikzlibrary{decorations}
\usetikzlibrary{decorations.pathmorphing,positioning}
\usetikzlibrary{decorations.markings}

\tikzset
{
hole/.style     = { draw = black, postaction = { decorate }, decoration = { markings, mark = at position .55 with { \arrow[black]{ triangle 45} } } },
spinwave_11/.style    = { draw = cyan, postaction = { decorate }, decoration = { markings, mark = at position .5 with { \arrow[cyan]{ triangle 45} } } },
particle_small_arrow/.style     = { draw = black, postaction = { decorate }, decoration = { markings, mark = at position .5 with { \arrow[scale = 0.6, black]{ triangle 45} } } },
spinwave_12/.style    = { draw = cyan, postaction = { decorate }, decoration = { markings, mark = at position .25 with { \arrow[cyan, >=triangle 45]{<} }, mark = at position .75 with { \arrow[cyan, >=triangle 45]{>} } } },
spinwave_21/.style    = { draw = cyan, postaction = { decorate }, decoration = { markings, mark = at position .25 with { \arrow[cyan, >=triangle 45]{>} }, mark = at position .75 with { \arrow[cyan, >=triangle 45]{<} } } },
spinwave_total/.style       = { decorate, decoration = {snake, amplitude = 2pt, segment length = 7pt } }
}

\tikzset{
  ncbar angle/.initial=90,
  ncbar/.style={
    to path=(\tikqtostart)
    -- ($(\tikqtostart)!#1!\pgfkeysvalueof{/tikz/ncbar angle}:(\tikqtotarget)$)
    -- ($(\tikqtotarget)!($(\tikqtostart)!#1!\pgfkeysvalueof{/tikz/ncbar angle}:(\tikqtotarget)$)!\pgfkeysvalueof{/tikz/ncbar angle}:(\tikqtostart)$)
    -- (\tikqtotarget)
  },
  ncbar/.default=0.5cm,
}

\usepackage{tgtermes} 

\newlength{\reducedsectionspacing}
\newlength{\defaultsectionspacing}
\newlength{\sectiontitlelength}

\newcommand{\shortsection}[2][]{
\section{\setlength{\sectiontitlelength}{\widthof{\thesection\quad}+\widthof{#2}}\ifdim\sectiontitlelength<\linewidth#2
\else
\fontdimen2\font=\reducedsectionspacing
\setlength{\sectiontitlelength}{\widthof{\thesection\quad}+\widthof{#2}}\ifdim\sectiontitlelength<\linewidth#2
\else\fontdimen2\font=\defaultsectionspacing
\ifx\\#1\\#2\else#1
\fi\fi\fi
}}

\tikzset{square left brace/.style={ncbar=0.1cm}}
\tikzset{square right brace/.style={ncbar=-0.1cm}}

\definecolor{myred}{RGB}{214,26,70}
\definecolor{myreddark}{RGB}{76,8,38}
\definecolor{myblue}{RGB}{35,106,185}
\definecolor{mybluedark}{RGB}{19,56,99}
\definecolor{mybluebright}{RGB}{225,236,249}

\def\te{{\rm e}}

\def\bd{{\bf d}}

\def\pa{\partial}
\def\nn{\nonumber}
\def\Re{{ \rm Re }}
\def\Im{{ \rm Im }}

\def\AF{{ \rm AF }}

\begin{document}
\title{Exact nonequilibrium hole dynamics, magnetic polarons and string excitations in antiferromagnetic Bethe lattices}
\date{\today}

\author{K. \ Knakkergaard \ Nielsen}
\affiliation{Max-Planck Institute for Quantum Optics, Hans-Kopfermann-Str. 1, D-85748 Garching, Germany}
\affiliation{Center for Complex Quantum Systems, Department of Physics and Astronomy, Aarhus University, Ny Munkegade 120, DK-8000 Aarhus C, Denmark}

\begin{abstract}
We investigate a rare instance of an exactly solvable nonequilibrium many-body problem. In particular, we derive an exact solution for the nonequilibrium dynamics of an initially localized single hole in a fully anisotropic antiferromagnetic Bethe lattice, described by the $t$-$J_z$ model. The solvability of the model relies on the fractal self-similarity of Bethe lattices, making it possible to compute the full motion of the hole as it moves through the lattice, as well as exactly characterizing the resulting effect on spin-spin correlation functions. We find that the hole remains bound to its initial position with large aperiodic oscillations in the hole density distribution. We track this back to the irregular pattern of the eigenenergies of the magnetic polaron ground state and string excitations, which we also determine exactly.
\end{abstract}

\maketitle

\section{Introduction}
\begin{figure}[t!]
\begin{center}
\includegraphics[width=1.0\columnwidth]{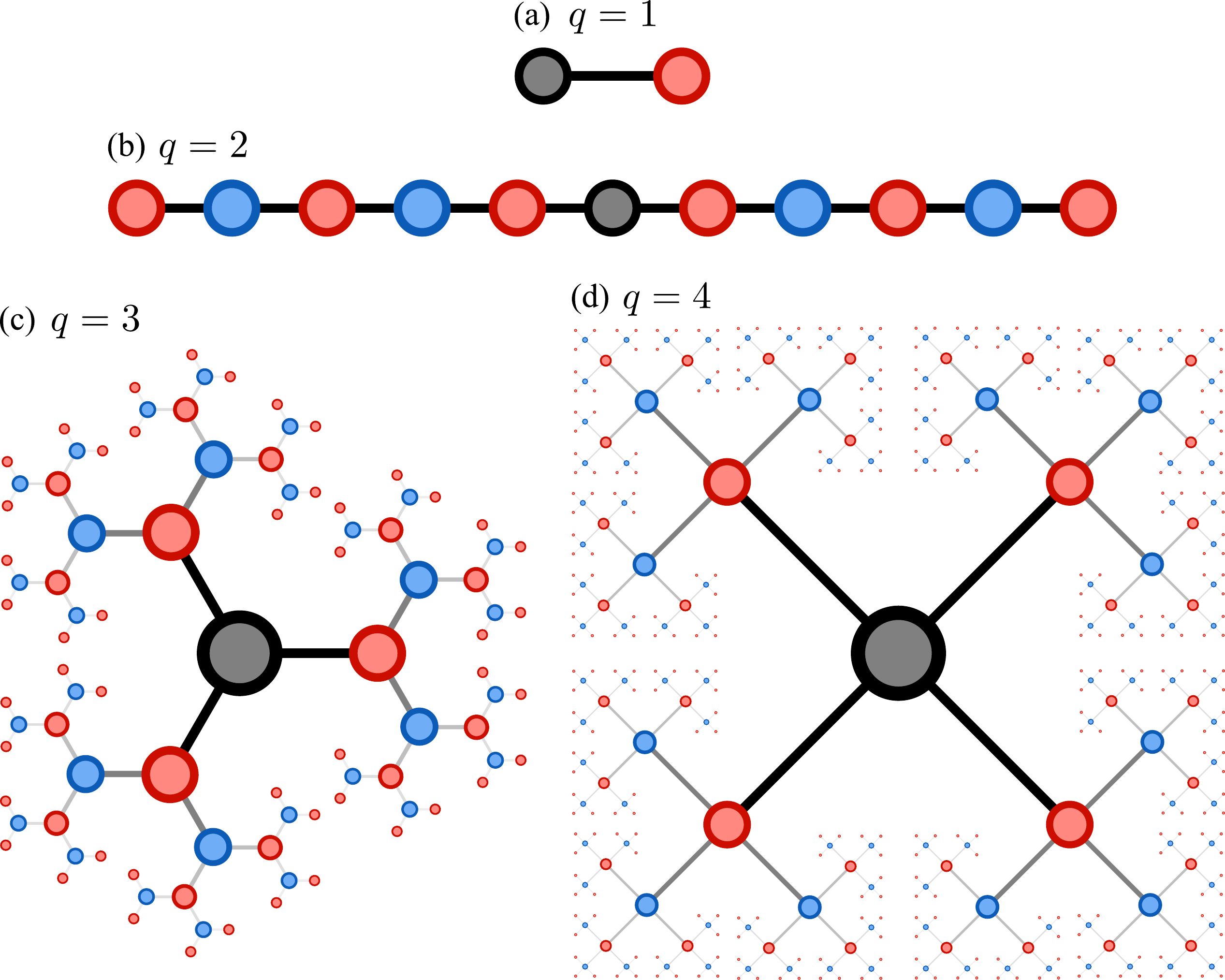}
\end{center}\vspace{-0.5cm}
\caption{\textbf{Single dopant in Bethe lattices}. A single hole (grey circle) is initialized at the center of an otherwise perfectly ordered antiferromagnetic Bethe lattice, showcased by coordination numbers $q = 1, 2, 3, 4$ nearest neighbors in panels (a) through (d). Here, red (blue) circles denote spin-$\uparrow$ (spin-$\downarrow$) fermions.}
\label{fig.bethe_lattices} 
\vspace{-0.25cm}
\end{figure}

The study of quantum many-body systems often rely on approximate descriptions such as mean-field \cite{Bogoliubov1947,Bardeen1957, BruusFlensberg} and variational treatments \cite{Griffiths2005,Landau_QM}, or on extensive numerical analyses such as quantum Monte Carlo \cite{Metropolis1949,Kalos1962,Hammond1994,Boninsegni2006,Kolorenc2011}. For this reason, instances in which the many-body dynamics can be solved exactly \cite{Katsura1974,Baxter1982,Andrei1995,Braak2011} are important to get precise insights into these systems, and may offer new ways to approximate related models . An important class of such systems are given by Bethe lattices [Fig. \ref{fig.bethe_lattices}], whose geometry is uniquely defined by the number of nearest neighbors. The name of these fascinating fractal structures are given in honor of the groundbreaking work of Hans Bethe on the Bethe ansatz \cite{Bethe1935}, being exact descriptions of e.g. ferro- and antiferromagnetism in these particular lattices \cite{Katsura1974}. Related studies have shown that the free quantum walk of a single particle in Bethe lattices may also be solved exactly \cite{Brinkman1970,Mahan2001,Eckstein2005,Economou2006,Nagy2017,Aryal2020}, which has been used to model amorphous solids \cite{Weaire1971,Weaire1971,Thorpe1971,Thorpe1973,Joannopoulos1974,Yndurain1975,Allan1980} and the hopping of ions in ice \cite{Chen1974}. 

%%%%%%%%%%%%%%%%%%%%%%%%%%%%%%%%%%%%%%%%%%%%%%%%%%%%%%%%%%%%%%%%%%% 
%%%%%%%%%%%%%%%%%%%%%%%%%%%%%%%%%%%%%%%%%%%%%%%%%%%%%%%%%%%%%%%%%%% 

In this article, we fuse these ideas to describe the motion of a single dopant in an antiferromagnetic Bethe lattice. In particular, we show that the nonequilibrium dynamics of a single hole hopping with amplitude $t$ in an antiferromagnetic environment with nearest neighbor spin-$z$ coupling $J_z$ can be solved exactly at zero temperature. Previously, an exact solution for the local hole Green's function, corresponding to the lowest order coefficient in the many-body wave function, was found \cite{Chernyshev1999,Wrzosek2021}. Here, we show that the full many-body wave function can be calculated, not only in the nonequilibrium case, but also for an important set of many-body eigenstates, i.e. the magnetic polaron ground state and the so-called string excitations \cite{Bulaevskii1968,Shraiman1988,Kane1989,Martinez1991,Liu1991,Reiter1994}. The appearance of magnetic polarons is a result of the inherent competition between the delocalization of the hole and the emergent magnetic frustrations. Such processes also give rise to induced interactions between two holes, which may provide a mechanism for high-temperature superconductivity \cite{Emery1987,Schrieffer1988,Dagotto1994}. Due to recent advances in quantum simulation using optical lattices \cite{Bakr2009,Esslinger2010,Sherson2010,Haller2015,Boll2016,Cheuk2016,Mazurenko2017,Hilker2017,Brown2017,Chiu2018,Brown2019,Koepsell2019,Chiu2019,Sanchez2020,Vijayan2020,Hartke2020,Sanchez2021,Ji2021,Gall2021,Yang2021}, systems in which these quasiparticles arise can now be realized and probed with unprecedented detail \cite{Koepsell2019,Chiu2019,Koepsell2020}. Consequently, magnetic polarons and the associated motion of holes in antiferromagnetic environments are also receiving renewed theoretical inquiries \cite{Carlstrom2016,Grusdt2018_2,Grusdt2018,Bohrdt2019,Bohrdt2020,Blomquist2020,Wang2021,Nielsen2021,Nielsen2022}. In particular, our present paper is related to the so-called string theory of magnetic polarons \cite{Grusdt2018_2}, in which the hole is effectively described as moving in a Bethe lattice. 

The exact solvability of the anisotropic $t$-$J_z$ model in Bethe lattices presented in this paper comes about as a result of the fractal self-similarity of the Bethe lattices. In particular, every time the hole hops, the system in the forward direction of motion looks the same as at the previous site. This underlying structure means that the wave function, expressed as a superposition of states with an increasing number of spin excitations, also attains a self-similar form, in which the coefficients of the wave function are related by quite simple recursion relations. This structure is very closely related to the approximate magnetic polarons states \cite{Reiter1994,Ramsak1998,Nielsen2021} and nonequilibrium wave function \cite{Nielsen2022} for the full $t$-$J$ model on square lattices. In fact, the present solution can also be understood in the context of the retraceable path approximation \cite{Brinkman1970}, which becomes exact in Bethe lattices. 

The article is organized as follows. Section \ref{sec.bethe_ising_antiferromagnet} describes the structure of the antiferromagnetic ground state, and the hopping Hamiltonian in the presence of holes. Section \ref{sec.holstein_primakoff} introduces the Holstein-Primakoff transformation, making it possible to give a concise description of the many-body states. Section \ref{sec.exact_solution} gives the derivation of the exact solution for the nonequilibrium many-body wave function. Section \ref{sec.eigenstates} describes the exact ground state and string excitations, before Sec. \ref{sec.results} characterizes the nonequilibrium hole and spin dynamics as a function of inverse interaction strength $J_z / t$ and coordination number $q$. 

\section{Ising antiferromagnets in Bethe lattices} \label{sec.bethe_ising_antiferromagnet}
We consider spin $1/2$ fermions hopping in Bethe lattice structures with $q$ nearest neighbors, as exemplified in Fig. \ref{fig.bethe_lattices}. We assume that the nearest neighbor spin-spin interactions are of the Ising type,
\begin{equation}
\hat{H}_J = J_z \sum_{\braket{{\bf i}, {\bf j}}} \left[\hat{S}^{(z)}_{\bf i}\hat{S}^{(z)}_{\bf j} - \frac{\hat{n}_{\bf i} \hat{n}_{\bf j}}{4} \right], 
\label{eq.H_J}
\end{equation}
and antiferromagnetic ($J_z > 0$). The Schwinger-fermion representation of spin $1/2$ as usual reads
\begin{equation}
\mathbf{S}_{\bf j}=\frac{1}{2}\sum_{\sigma,\sigma'}\hat{c}_{{\bf j},\sigma}^\dagger\boldsymbol{\sigma}_{\sigma\sigma'}\hat{c}_{{\bf j},\sigma'}
\end{equation}
with $\boldsymbol{\sigma}=(\sigma_x,\sigma_y,\sigma_z)$ a vector of the Pauli matrices. The antiferromagnetic ground state is, thus, achieved by having every neighboring spin pointing in opposite directions. Choosing a specific lattice site as the central reference point, we then speak of the depth $d$ of a given site, as the number of hops it takes to go to that site. Correspondingly, the total depth $d_{\rm tot}$ of the lattice is the maximum number of jumps that can be made from the central site. In Fig. \ref{fig.bethe_lattices}(c) and (d), the cases of $q = 3$ and $q = 4$ respectively, the total depth of the shown lattices is $d_{\rm tot} = 5$. The total number of sites in a Bethe lattice with $q$ nearest neighbors and total depth $d_{\rm tot}$ is
\begin{equation}
N(q,d_{\rm tot}) = 1 + q \sum_{j = 0}^{d_{\rm tot} - 1} (q - 1)^j = 1 + \frac{(q - 1)^{d_{\rm tot}} - 1}{q - 2}.
\label{eq.number_of_sites} 
\end{equation}
In the case of $q = 2$, the Bethe lattice collapses to a one-dimensional chain [Fig. \ref{fig.bethe_lattices}(b)], and the expression $N = 1 + d_{\rm tot}$ can be found from Eq. \eqref{eq.number_of_sites} by applying l'Hospital's rule to the limit $q \to 2$. Generally, the number of nearest neighbor links is simply $N - 1$, and the antiferromagnetic energy contribution from each of these is $-J_z/2$. Therefore, the ground state energy for an Ising antiferromagnet in a Bethe lattice is simply
\begin{equation}
E_0(q,d_{\rm tot}) = - \left[N(q,d_{\rm tot}) - 1\right]\frac{J_z}{2}. 
\label{eq.ground_state_energy_antiferromagnet}
\end{equation}
Since the number of sites at a depth $d + 1$ is $q - 1$ times larger than the number of sites at depth $d$, and since the spins at $d+1$ point opposite to the spins at $d$, the lattice has a nonzero total spin
\begin{align}
S^{(z)}_{\rm tot}(q, d_{\rm tot}) &= -\frac{1}{2}\left[1 - q\sum_{j = 0}^{d_{\rm tot} - 1} (1 - q)^j \right] \nn \\
&= -\frac{1}{2}(1-q)^{d_{\rm tot}},
\label{eq.S_z_tot_ground_state}
\end{align}
assuming that the central site has $S^{(z)} = -1 / 2$. This characterizes the underlying antiferromagnetic ground state. Below half-filling, we assume that the fermionic particles can hop between nearest neighbor sites with amplitude $t$
\begin{equation}
\hat{H}_t = - t\sum_{\braket{{\bf i}, {\bf j}},\sigma} \left[\tilde{c}^\dagger_{{\bf i}, \sigma}\tilde{c}_{{\bf j}, \sigma} + {\rm H.c.} \right], 
\label{eq.H_t}
\end{equation}
where $\tilde{c}^\dagger_{{\bf j},\sigma} = \hat{c}^\dagger_{{\bf j},\sigma}(1 - n_{\bf j})$, and the factor $1 - n_{\bf j}$ restrains the Hilbert space to states with maximally one fermion per lattice site.

\section{The Holstein-Primakoff transformation} \label{sec.holstein_primakoff}
The Holstein-Primakoff transformation is performed to give a more efficient description of the system just below half-filling, in terms of bosonic spin excitation operators $\hat{s}_{\bf i}$, and fermionic hole operators $\hat{h}_{\bf i}$. To perform the transformation, it is first useful to define two sublattices corresponding to the sites that in the absence of holes host spins pointing up and down respectively. More precisely, fermions on every site at an odd-numbered depth, $d = 1, 3, 5, \dots$, initially has spin-$\uparrow$, and is defined to lie on sublattice A. Similarly, every site at an even-numbered depth, $d = 0, 2, 4, \dots$, belong to sublattice B, having spin-$\downarrow$ fermions. We may, then, define the Holstein-Primakoff transformation according to
\begin{align}
{\rm {\bf A}}: \hat{S}^{-}_{\bf i} &= \hat{s}_{\bf i}^\dagger F(\hat{h}_{\bf i}, \hat{s}_{\bf i}), \; \tilde{c}_{{\bf i},\downarrow}=\hat{h}^\dagger_{\bf i}\hat{S}_{\bf i}^{+}, \nn \\
\tilde{c}_{{\bf i},\uparrow} &=\hat h^\dagger_{\bf i} F(\hat{h}_{\bf i}, \hat{s}_{\bf i}), \; \hat{S}_{\bf i}^z= \left[\frac{1}{2} - \hat{s}^\dagger_{\bf i}\hat{s}_{\bf i}\right] [1-\hat{h}_{\bf i}^\dagger \hat{h}_{\bf i}]. \nn \\
{\rm {\bf B}}: \hat{S}^{+}_{\bf j} &= \hat{s}_{\bf j}^\dagger F(\hat{h}_{\bf j}, \hat{s}_{\bf j}), \; \tilde{c}_{{\bf j},\uparrow}=\hat{h}^\dagger_{\bf j}\hat{S}_{\bf j}^{-}, \nn \\
\tilde{c}_{{\bf j},\downarrow} &= \hat h^\dagger_{\bf j} F(\hat{h}_{\bf j}, \hat{s}_{\bf j}), \; \hat{S}_{\bf j}^z= \left[\hat{s}^\dagger_{\bf j}\hat{s}_{\bf j} - \frac{1}{2}\right] [1-\hat{h}_{\bf j}^\dagger \hat{h}_{\bf j}],
\end{align}
with $F(\hat{s},\hat{h}) = \sqrt{1 - \hat{s}^\dagger \hat{s} - \hat{h}^\dagger \hat{h}}$. Consequently, the spin-spin interaction is rewritten as 
\begin{align}
\hat{H}_J = -J_z \sum_{\braket{{\bf i}, {\bf j}}} & [1-\hat{h}_{\bf i}^\dagger \hat{h}_{\bf i}] \left[\left(\frac{1}{2} - \hat{s}^\dagger_{\bf i}\hat{s}_{\bf i}\right)\left(\frac{1}{2} - \hat{s}^\dagger_{\bf j}\hat{s}_{\bf j}\right) + \frac{1}{4} \right] \nn \\
\cdot &[1-\hat{h}_{\bf j}^\dagger \hat{h}_{\bf j}]. 
\label{eq.H_J_holstein_primakoff}
\end{align}
This expression fully accounts for the presence of holes, in which case the nearest neighbor spin coupling is naturally $0$. Similarly, the hopping Hamiltonian becomes
\begin{align}
\hat{H}_t = t \sum_{\braket{{\bf i}, {\bf j}}} \! F(\hat{s}_{\bf i})F(\hat{s}_{\bf j})  \left[ \hat{h}^\dagger_{\bf j}\hat{h}_{\bf i} \hat{s}_{\bf j} + \hat{h}^\dagger_{\bf i}\hat{h}_{\bf j} \hat{s}_{\bf i} \right] + {\rm H.c.}
\label{eq.H_t_holstein_primakoff}
\end{align}
Here, due to the fermionic statistics of the holes, we may use that $F(\hat{s}_{\bf i},\hat{h}_{\bf i}) \hat{h}_{\bf i} = F(\hat{s}_{\bf i},0) \hat{h}_{\bf i}$. Additionally, we simplify the notation by writing $F(\hat{s}) = F(\hat{s},0) = \sqrt{1 - \hat{s}^\dagger\hat{s}}$. Equation \eqref{eq.H_t_holstein_primakoff} shows that as the hole hops through the lattice, it may do so by absorbing or emitting spin excitations. The factors of $F(\hat{s}_{\bf i})F(\hat{s}_{\bf j})$ constrains this process, so that there is always either $0$ or $1$ spin excitation on each site, yielding the exact hardcore constraint.  

\section{The exact nonequilibrium wave function} \label{sec.exact_solution}
We are now ready to efficiently formulate the exact nonequilibrium wave function. We will focus on Bethe lattices with more than 1 nearest neighbors, $q\geq 2$, as the case of $q = 1$ corresponds to a system of only two sites. We assume that the hole is initially located at the central site of the Bethe lattice. Note that in the thermodynamic limit, $d_{\rm tot} \to \infty$, any site in the lattice is equivalent. Therefore, this choice for the initial state is actually the general starting point for quenching the system from the antiferromagnet ground state into a state with a single localized hole. A lattice site at depth $d$ is written as ${\bf j}_d = 0, j_1, \dots, j_d$. Here, $j_1 = 1, 2, \dots, q$, and $j_l = 1, 2, \dots, q-1$ for $l\geq 2$ describe the sites at each depth, similar to Ref. \cite{Katsura1974}. In this manner, ${\bf j}_d$ gives the full path from the central site ${\bf j}_0 = 0$ to the specific site of interest at depth $d$. Two sites ${\bf j}_d$ and ${\bf l}_{d+1}$ at depth $d$ and $d+1$ are, thus, nearest neighbors only if ${\bf l}_{d+1} = {\bf j}_d, l_{d+1}$. This construction allows us to write the full nonequilibrium wave function at time $\tau$ as
\begin{align}
&\!\!\! \ket{\Psi(\tau)} = C^{(0)}(\tau) \cdot \hat{h}^{\dagger}_{0} \ket{\AF} + C^{(1)}(\tau) \sum_{{\bf j}_1} \hat{h}^\dagger_{{\bf j}_1} \hat{s}^\dagger_{0} \ket{\AF} \nn \\
&\phantom{\ket{\Psi(\tau)} =} + C^{(2)}(\tau) \sum_{{\bf j}_2} \hat{h}^\dagger_{{\bf j}_2} \hat{s}^\dagger_{0}\hat{s}^\dagger_{{\bf j}_1} \ket{\AF} + \dots \nn \\
&\!\!\! = C^{(0)}(\tau) \hat{h}^{\dagger}_{0} \ket{\AF} + \sum_{d = 1}^{d_{\rm tot}} C^{(d)}(\tau) \sum_{{\bf j}_d} \hat{h}^\dagger_{{\bf j}_d} \prod_{l = 0}^{d-1} \hat{s}^\dagger_{{\bf j}_l} \ket{\AF}. \!
\label{eq.exact_many_body_wave_function}
\end{align}
In this expression, the lattice site at depth $d$ is always constrained to be next to the one at depth $d - 1$: ${\bf j}_d = {\bf j}_{d-1}, j_d$. Note that the coefficients of the wave function $C^{(d)}$ \emph{cannot} depend on the exact path to the lattice point, ${\bf j}_d = j_1, \dots, j_d$, only the depth $d$ of that point. The underlying reason is that the system is symmetric in all these paths, and that the initial wave function, $\ket{\Psi(\tau = 0)} = \hat{h}^{\dagger}_{0} \ket{\AF}$, is as well. This gives a remarkable simplification of the treatment, as it describes the dynamics in a system whose size grows exponentially with the depth [see Eq. \eqref{eq.number_of_sites}], in terms of a linear number of coefficients $C^{(0)}, \dots, C^{(d_{\rm tot})}$.

\subsection{General time-dependent solution} \label{subsec.general_solution}
To progress on solving for the coefficients of the nonequilbrium wave function, we introduce the retarded and advanced wave functions
\begin{align}
\ket{\Psi(\tau)} &= \ket{\Psi^{\rm R} (\tau)} + \ket{\Psi^{\rm A}(\tau)} \nn \\
&= \te^{-\eta|\tau|}\left[\theta(\tau)\ket{\Psi(\tau)} + \theta(-\tau)\ket{\Psi(\tau)}\right],
\label{eq.retarded_and_advanced_state}
\end{align}
allowing us to regularize the Fourier transformation to frequency space with the positive infinitesimal $\eta > 0$, and the Heaviside step function $\theta(\tau)$. Consequently, we may express the Schr{\"o}dinger equation, $i\pa_{\tau} \ket{\Psi(\tau)} = \hat{H}\ket{\Psi(\tau)}$, as 
\begin{align}
(\omega + i\eta) \ket{\Psi^{\rm R}(\omega)} &= +i\ket{\Psi(\tau = 0)} + \hat{H} \ket{\Psi^{\rm R}(\omega)}, \nn \\
(\omega - i\eta) \ket{\Psi^{\rm A}(\omega)} &= -i\ket{\Psi(\tau = 0)} + \hat{H} \ket{\Psi^{\rm A}(\omega)}.
\label{eq.schrodinger_equation_frequency_space}
\end{align}
As these equations are complex conjugate of each other, it follows that $\ket{\Psi^{\rm A}(\omega)} = [\ket{\Psi^{\rm R}(\omega)}]^*$. We denote the coefficients of the retarded state $R^{(d)}(\omega)$. The dynamical coefficients in Eq. \eqref{eq.exact_many_body_wave_function} are then retrieved from $R^{(d)}(\omega)$ by a Fourier transformation
\begin{equation}
C^{(d)}(\tau) = \int \frac{{\rm d}\omega}{2\pi} \te^{-i(\omega + i\eta)\tau} \cdot 2\Re[R^{(d)}(\omega)]. 
\label{eq.C_d_and_R_d}
\end{equation}
Using Eq. \eqref{eq.schrodinger_equation_frequency_space}, the equations of motion for the retarded state coefficients become
\begin{align}
\!\!\!\!\!\!\!\!(\omega + i\eta) R^{(0)}(\omega) &= i + E_J^{(0)} R^{(0)}(\omega) + qt R^{(1)}(\omega), \!\!\!\nn \\
\!\!\!\!\!\!\!\!(\omega + i\eta) R^{(d)}(\omega) &= E_J^{(d)} R^{(d)}(\omega) + t R^{(d-1)}(\omega) \!\!\!\nn \\
\!\!\!\!\!\!\!\!&\phantom{=} + (q-1) t R^{(d+1)}(\omega), \!\!\!\nn \\
\!\!\!\!\!\!\!\!(\omega + i\eta) R^{(d_{\rm tot})}(\omega) &= E_J^{(d_{\rm tot})} R^{(d_{\rm tot})}(\omega) + t R^{(d_{\rm tot}-1)}(\omega). \!\!\!
\label{eq.equations_of_motion}
\end{align}
The term $i$ in the equation for $R^{(0)}(\omega)$ comes directly from the term $i\ket{\Psi(\tau = 0)}$ in Eq. \eqref{eq.schrodinger_equation_frequency_space}. Furthermore, the magnetic energy cost $E_J^{(d)}$ comes from applying the spin-spin interaction Hamiltonian in Eq. \eqref{eq.H_J_holstein_primakoff} to the $d$th term in the wave function. Finally, the hopping Hamiltonian simply relates the coefficient at depth $d$ to the coefficients at depth $d - 1$ and $d + 1$. At $d = 0$, there are $q$ nearest neighbors at depth $d = 1$. All subsequent depths $d \leq d_{\rm tot} - 1$, however, only has $q-1$ nearest neighbors at depth $d+1$ and a single neighbor one depth up, $d - 1$. This leads to the terms $(q-1) \cdot t \cdot R^{(d + 1)}(\omega)$ and $1 \cdot t \cdot R^{(d-1)}(\omega)$ for $1 \leq d \leq d_{\rm tot}$. Finally, by assumption the bottom depth is $d_{\rm tot}$, and so $R^{(d)} = 0$ for $d > d_{\rm tot}$.

%%%%%%%%%%%%%%%%%%%%%%%%%%%%%%%%%%%%%%%%%%%%%%%%%%%%%%%%%%%%%%%%%%% 
\begin{figure}[t!]
\begin{center}
\includegraphics[width=0.9\columnwidth]{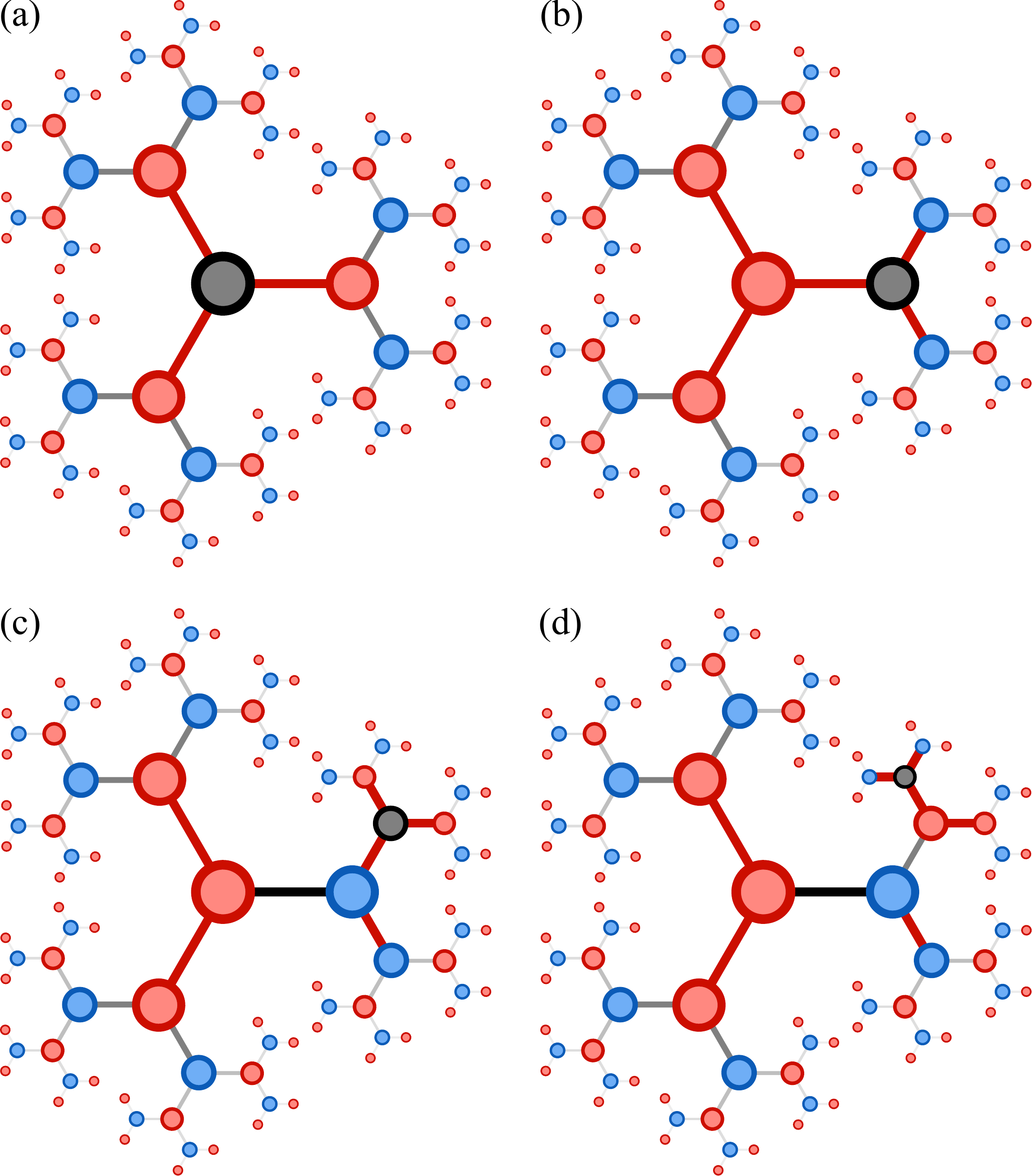}
\end{center}\vspace{-0.5cm}
\caption{\textbf{Broken antiferromagnetic spin bonds}. As the hole moves into the Bethe lattice [(a)--(d)], a number of antiferromagnetic spin bonds are broken along the way (red lines), illustrated here in the case of $q = 3$. (a) At $d = 0$, there are $q = 3$ broken bonds, giving a spin energy of $E_J^{(0)} = E_0 + qJ_z / 2$. (b) For $d = 1$, there are $q - 1 = 2$ further broken bonds, so $E_J^{(1)} = E_J^{(0)} + (q-1)J/2$. (c) and (d) For $d\geq 2$, the spin bonds along the path of the hole are repaired. As a result, there are $q-1$ new spin bonds broken and $1$ repaired, yielding $q - 2 = 1$ in the present case. Therefore, $E_J^{(d+1)} = E_J^{(d)} + (q-2)J_z/2$ for $d\geq 1$.}
\label{fig.bethe_lattices_broken_bonds} 
\vspace{-0.25cm}
\end{figure} 
%%%%%%%%%%%%%%%%%%%%%%%%%%%%%%%%%%%%%%%%%%%%%%%%%%%%%%%%%%%%%%%%%%% 

The value of $E_J^{(d)}$ may be understood recursively, starting from $d = 0$. Here, the hole is located at the central site, and there are $q$ broken antiferromagnetic spin bonds of strength $J_z / 2$, as indicated in Fig. \ref{fig.bethe_lattices_broken_bonds}(a). Note that Eq. \eqref{eq.H_J_holstein_primakoff} implies that a hole at a site has the same magnetic energy cost as a spin excitation at that site. Therefore, $E_J^{(0)} = E_0(q,d) + qJ_z / 2$, where $E_0(q,d)$ is the ground state energy of the Ising antiferromagnet [Eq. \eqref{eq.ground_state_energy_antiferromagnet}]. When the hole is at depth $d = 1$, as shown in Fig. \ref{fig.bethe_lattices_broken_bonds}(b), there are $q-1$ additional broken bonds. So $E_J^{(1)} = E_J^{(0)} + (q-1)J_z / 2$. At every subsequent depth $d\geq 2$ [Figs. \ref{fig.bethe_lattices_broken_bonds}(c) and \ref{fig.bethe_lattices_broken_bonds}(d)], the spin bonds along the path of the hole are repaired. Therefore, each hop is associated with $q-1$ new broken bonds and $1$ repaired bond. The magnetic energy cost for $d\geq 2$ is, thus, $(q-2) J_z / 2$ for every step in depth. In this way, 
\begin{align}
E_J^{(d)} &= \left[(q-1) + (d-1) (q-2)\right]\frac{J_z}{2},
\label{eq.E_J}
\end{align}
for $1\leq d \leq d_{\rm tot} - 1$. Here, we let $E_J^{(d)} \to E_J^{(d)} - E_J^{(0)}$, measuring all energies with respect to $E_J^{(0)}$. Finally, at depth $d_{\rm tot}$ the absence of neighbors at $d_{\rm tot} + 1$ in principle results in a surface effect, in which there is no additional magnetic energy cost for the hole to hop to the surface, $E_J^{(d_{\rm tot})} = E_J^{(d_{\rm tot} - 1)}$. While this presumably has an effect on the dynamics for small lattices, we focus on the behavior for large total depths, $d_{\rm tot} \gg 1$, and ignore the surface effects. With the magnetic energy costs for the hole to hop to the $d$th depth in place [Eq. \eqref{eq.E_J}], we can now solve the equations of motion in Eq. \eqref{eq.equations_of_motion} iteratively. Specifically, the equation for $R^{(d_{\rm tot})}$ simply yields
\begin{align}
R^{(d_{\rm tot})}(\omega) &= \frac{t}{\omega + i\eta - E_J^{(d_{\rm tot})}} R^{(d_{\rm tot} - 1)}(\omega) \nn \\
&= t \cdot G^{(d_{\rm tot})}(\omega) \cdot R^{(d_{\rm tot} - 1)}(\omega).
\label{eq.R_d_tot_omega}
\end{align}
Continuing this process for $d_{\rm tot} \geq d \geq 1$, we get the recursion relation
\begin{align}
R^{(d)}(\omega) &= t \cdot G^{(d)}(\omega) \cdot R^{(d - 1)}(\omega), \nn \\
G^{(d)}(\omega) &= \frac{1}{\omega + i\eta - E_J^{(d)} - (q-1)t^2 \cdot G^{(d+1)}(\omega)},
\label{eq.recursion_relation}
\end{align}
defining $G^{(d_{\rm tot}+1)}(\omega) = 0$. We refer to $G^{(d)}(\omega)$ as the retarded Green's function at depth $d$. Finally, inserting the solution from $d = 1$ into the equation for $R^{(0)}$, we get $R^{(0)}(\omega) = iG^{(0)}(\omega)$, with 
\begin{equation}
G^{(0)}(\omega) = \frac{1}{\omega + i\eta - q t^2 \cdot G^{(1)}(\omega)}.
\label{eq.G_0}
\end{equation}
Using the recursion relation in Eq. \eqref{eq.recursion_relation} in reverse, we obtain the general result for the coefficients of the retarded wave function in frequency space
\begin{align}
R^{(0)}(\omega) &= iG^{(0)}(\omega), \nn \\
R^{(d)}(\omega) &= iG^{(0)}(\omega) \cdot t^{d} \cdot \prod_{l = 1}^d G^{(l)}(\omega), \; 1 \leq d \leq d_{\rm tot}.
\label{eq.R_d_general}
\end{align}
We end this subsection by connecting the coefficients of the wave function to a specific set of many-body correlation functions. Specifically, we have that
\begin{align}
R^{(0)}(\tau) &= \theta(\tau) \bra{\AF}\hat{h}_0 \te^{-iH\tau} \hat{h}_0^\dagger \ket{\AF}, \nn \\
R^{(d)}(\tau) &= \theta(\tau) \bra{\AF}\hat{h}_{{\bf j}_d} \prod_{l=0}^{d-1} \hat{s}_{{\bf j}_l} \te^{-iH\tau} \hat{h}_0^\dagger \ket{\AF}.
\label{eq.correlation_functions}
\end{align}
This more clearly demonstrates that the $d$th coefficient of the wave function is the probability amplitude of the hole to be at depth $d$ at site ${\bf j}_d$ with $d$ spin excitations after a time $\tau$.

\subsection{Thermodynamic limit} \label{subsec.thermodynamic_limit}
In the thermodynamic limit, $d_{\rm tot}\to\infty$, the recursion relation for the retarded Green's functions in Eq. \eqref{eq.recursion_relation} can be written as a self-consistency equation for a single Green's function $G_1(\omega)$. Specifically, we can write $G^{(d)}(\omega) = G_1(\omega - E_J^{(d)})$ for $d\geq 1$, where
\begin{equation}
G_1(\omega) = \frac{1}{\omega - (q-1)t^2\, G_1(\omega - (q-2)\frac{J_z}{2})}.\!
\label{eq.G_1_self_consistency}
\end{equation}
For brevity, in this subsection we let $\omega + i\eta\to\omega$. Similarly, we get the Green's function for the central site, $d = 0$, to be
\begin{equation}
G_0(\omega) = \frac{1}{\omega - qt^2 \, G_1(\omega - (q-1)\frac{J_z}{2})}.
\label{eq.G_0_self_consistency}
\end{equation}
Inspired by Ref. \cite{Chernyshev1999}, we anticipate that the $G_1$ Green's function can be expressed in terms of Bessel functions. In fact, using the ansatz $G_1(\omega) = -1/(\sqrt{q-1}t) \cdot Y(\omega) / Y(\omega + (q-2)J/2)$, the resulting recursive relation for the $Y$ functions coincide with that for the Bessel functions of the first kind. This results in
\begin{equation}
G_1(\omega) = -\frac{1}{\sqrt{q-1}t} \frac{J_{\Omega(\omega)}\left(\frac{4\sqrt{q-1}t}{(q-2)J_z}\right)}{J_{\Omega(\omega) - 1}\left(\frac{4\sqrt{q-1}t}{(q-2)J_z}\right)},
\label{eq.G_1_solution}
\end{equation}
for $q\geq 3$, with $\Omega(\omega) = -2\omega / (q-2)J_z$. This generalizes the result in Ref. \cite{Chernyshev1999}, where the $q = 4$ Bethe lattice is effectively used as an approximate description of a two-dimensional square lattice, and coincides with what is found in Ref. \cite{Wrzosek2021}. We can now use the simple fraction of Bessel functions in Eq. \eqref{eq.G_1_solution} together with $G^{(d)}(\omega) = G_1(q, \omega - E_J^{(d)})$ for $d\geq 1$, to find the explicit expression for the coefficients of the nonequilibrium many-body wave function in the thermodynamic limit
\begin{equation}
R^{(d)}(\omega) = \frac{iG_0(\omega)}{(-\sqrt{q-1})^d}\cdot \frac{J_{\Omega(\omega - E_J^{(d)})}\left(\frac{4\sqrt{q-1}t}{(q-2)J_z}\right)}{J_{\Omega(\omega - E_J^{(1)}) - 1}\left(\frac{4\sqrt{q-1}t}{(q-2)J_z}\right)},
\label{eq.R_d_thermodynamic_limit}
\end{equation}
where the magnetic energy cost for a hole at depth $d$, $E_J^{(d)}$, is given in Eq. \eqref{eq.E_J}. In the limit of infinitely strong interactions, $J_z / t \to 0^+$, Eq. \eqref{eq.G_1_self_consistency} leads to a second order equation for $G_1$, yielding
\begin{gather}
G_1(\omega) \!\overset{J \to 0^+}{=}\! \frac{\omega}{2(q-1)t^2} \left(1 - \sqrt{1 - \frac{4(q-1)t^2}{\omega^2}}\right).
\label{eq.G_1_J_0_limit}
\end{gather}
This results in the hole Green's function
\begin{gather}
G_0(\omega) \overset{J \to 0^+}{=} \! \frac{1}{\omega \left[1 - \frac{q}{2(q-1)}\left(1 - \sqrt{1-\frac{4(q-1)t^2}{\omega^2}}\right)\right]}.
\label{eq.J_0_limit}
\end{gather}
Equations \eqref{eq.G_1_J_0_limit} and \eqref{eq.J_0_limit} together with Eq. \eqref{eq.R_d_general} describes a free quantum walk of a single particle in Bethe lattices \cite{Eckstein2005}, as one might expect in this extreme limit, where spin interactions play no role. 

\section{Exact ground state and string excitations} \label{sec.eigenstates}
The present methodology allows us to extract a certain set of many-body eigenstates on top of the nonequilibrium wave function derived in Sec. \ref{subsec.general_solution}. This constitutes the \emph{depth symmetric} states
\begin{equation}
\ket{\Psi_n} = C^{(0)}_n \hat{h}^{\dagger}_{0} \ket{\AF} + \sum_{d = 1}^{d_{\rm tot}} C^{(d)}_n \sum_{{\bf j}_d} \hat{h}^\dagger_{{\bf j}_d} \prod_{l = 0}^{d-1} \hat{s}^\dagger_{{\bf j}_l} \ket{\AF},
\label{eq.Psi_n}
\end{equation} 
which are all $d_{\rm tot} + 1$ states for which the coefficients of the wave function only depend on the depth $d$ of the hole, and not the particular point ${\bf j}_d$ at which it is located. Put in another manner, they are all the states that can be reached dynamically by immersing a single hole at the center of the lattice, i.e. the ones for which $\braket{\Psi_n|\Psi(\tau)} \neq 0$. In the limit of weak interactions, $J_z \gg t$, the many-body ground state is just the lowest order term, $\hat{h}^{\dagger}_{0} \ket{\AF}$, as the hopping of the hole is strongly suppressed. Conversely, in the limit of infinite interactions, $J_z / t \to 0^+$, the energy should go to that of a free particle, which in a Bethe lattice is $-2\sqrt{q-1}t$. This is indeed the case for our solution, shown explicitly in Sec. \ref{subsec.strong_interaction_limit}. Therefore, we expect $\ket{\Psi_0}$ to be the true many-body ground state for any value of $J_z / t \geq 0$. This is in contrast to the case of regular lattices, in which this type of states, so-called string excitations, do not approach the correct limiting value at very small values of $J_z / t$. Instead, as a precursor to the Nagaoka limit at $J = 0$ \cite{Nagaoka1966}, a ferromagnetic polaron is expected to emerge \cite{White2001}.

\subsection{General time-independent solution} \label{subsec.many_body_eigenstates}
The static Schr{\"o}dinger equation,
\begin{equation}
\varepsilon_n \ket{\Psi_n} = \hat{H} \ket{\Psi_n}
\end{equation}
simply corresponds to removing the $i\ket{\Psi(\tau = 0)}$ term in Eq. \eqref{eq.schrodinger_equation_frequency_space}, and replacing the frequency $\omega + i\eta$ with the eigenstate energy $\varepsilon_n$. As a result, we simply have to remove the term $i$ in the equations of motion in Eq. \eqref{eq.equations_of_motion} for the nonequilibrium case, and the same recursive relation in Eq. \eqref{eq.recursion_relation} for the nonequilibrium coefficients, therefore, applies to these eigenstates. Explicitly,
\begin{equation}
C^{(d)}_n = t \cdot G^{(d)}(\varepsilon_n) \cdot C^{(d - 1)}_n,
\label{eq.recursion_relation_eigenstate}
\end{equation}
with the Green's functions $G^{(d)}$ evaluated at the quasiparticle pole $\varepsilon_n$. Inserting the result at depth $d = 1$ into the equation of motion at depth $d = 0$, we get
\begin{equation}
\varepsilon_n \cdot C^{(0)}_n = qt^2 \cdot G^{(1)}(\varepsilon_n) \cdot C^{(0)}_n \Rightarrow \varepsilon_n = \Sigma^{(0)}(\varepsilon_n). 
\label{eq.equation_for_epsilon_n}
\end{equation}

%%%%%%%%%%%%%%%%%%%%%%%%%%%%%%%%%%%%%%%%%%%%%%%%%%%%%%%%%%%%%%%%%%% 
\begin{figure}[t!]
\begin{center}
\includegraphics[width=1.0\columnwidth]{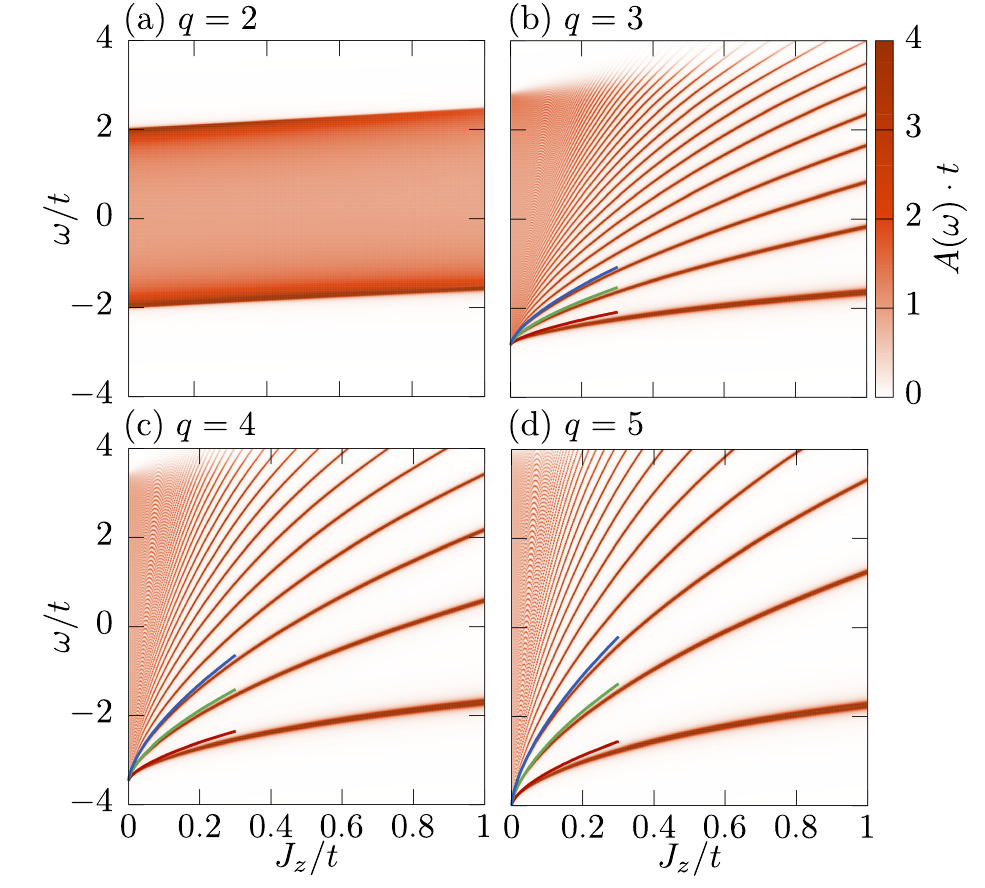}
\end{center}\vspace{-0.5cm}
\caption{\textbf{Spectral functions}, $A(\omega)$, for 4 indicated values of the number of nearest neighbors, $q$. (a) For $q = 2$, the particles sit in a 1D chain [Fig. \ref{fig.bethe_lattices}(b)], and there is a continuous spectrum above the quasiparticle pole at energy $\varepsilon_0 = J_z / 2 - \sqrt{(J_z / 2)^2 + (2t)^2}$. For higher values of $q$ [(b)--(d)], the spectrum consists of a series quasiparticle peaks, scaling with $(J_z / t)^{2/3}$ at strong interactions (colored lines).  }
\label{fig.spectral_functions} 
\vspace{-0.25cm}
\end{figure} 
%%%%%%%%%%%%%%%%%%%%%%%%%%%%%%%%%%%%%%%%%%%%%%%%%%%%%%%%%%%%%%%%%%% 

Here, $\Sigma^{(0)}(\omega) = qt^2 \cdot G^{(1)}(\omega)$ is the self-energy associated with the Green's function $G^{(0)}(\omega)$ in Eq. \eqref{eq.G_0}. This equation simply states the fact that the poles of $G^{(0)}$ are eigenstate energies of the system, the lowest of which gives the true ground state of the system. To find the lowest order coefficient of the $n$th state, $C^{(0)}_n$, we compute the full norm of the wave function, $\ket{\Psi_n}$. Repeated use of the recursion relation in Eq. \eqref{eq.recursion_relation_eigenstate} yields
\begin{align}
&1 = \braket{\Psi_n|\Psi_n} = \left|C^{(0)}_n\right|^2 + q \sum_{d = 1}^{d_{\rm tot}} (q-1)^{d-1} \left|C^{(d)}_n\right|^2 \nn \\
  &= \left|C^{(0)}_n\right|^2 \Big[1 + qt^2 \big(G^{(1)}(\varepsilon_n)\big)^2 \nn \\
  &\phantom{=\left|C^{(0)}_n\right|^2 \Big[} + q(q-1)t^4 \big(G^{(1)}(\varepsilon_n)G^{(2)}(\varepsilon_n)\big)^2 + \dots \Big] \nn \\
  &= \left|C^{(0)}_n\right|^2\! \Big[1 + qt^2 \big(G^{(1)}(\varepsilon_n)\big)^2 \big[1 + (q-1)t^2 G^{(2)}[1 + \dots] \big] \Big] \nn
\end{align} 
Investigating the quasiparticle residue, $Z_n = [1 - \pa_{\omega}\Sigma^{(0)}(\omega)|_{\omega = \varepsilon_n}]^{-1}$, shows that the expression within the square brackets above is simply $Z^{-1}_n$. In turn, $C^{(0)}_n = \sqrt{Z_n}$, whereby we may write the eigenstate coefficients as
\begin{align}
C^{(0)}_n &= \sqrt{Z_n}, \nn \\
C^{(d)}_n &= \sqrt{Z_n} \cdot t^{d} \cdot \prod_{l = 1}^d G^{(l)}(\varepsilon_n), \; 1 \leq d \leq d_{\rm tot}.
\label{eq.C_d_n_general}
\end{align}
In the thermodynamic limit, these coefficients can, in complete analogy to the nonequilibrium case investigated in Sec. \ref{subsec.thermodynamic_limit}, also be written as fractions of Bessel functions. This has previously been established for the ground state in Ref. \cite{Bieniasz2019}, where the Bethe lattice description is used as a variational ansatz for the square lattice case. In Fig. \ref{fig.spectral_functions}, we plot the spectral function $A(\omega) = -2\Im\, G_0(\omega)$, computed in the thermodynamic limit from Eq. \eqref{eq.G_0_self_consistency}. This reveals the poles of $G_0$, and thereby the eigenstate energies $\varepsilon_n$. In the special case of $q = 2$ \cite{Batista2000}, the system is one-dimensional and the spectral function splits into a quasiparticle pole at
\begin{equation}
\varepsilon_0 = \frac{J_z}{2} - \sqrt{4t^2 + \frac{J_z^2}{4}}. 
\label{eq.epsilon_0_q_2}
\end{equation}
and a continuum of states residing between $J_z/2 \pm 2t$; see also Fig. \ref{fig.spectral_functions}(a). This describes the effect known as spin-charge separation, which has been intensely studied both theoretically and experimentally in one-dimensional systems \cite{Tomonaga1950,Kim1996,Segovia1999,Auslaender2005,Kim2006,Jompol2009,Vijayan2020}. In the present case, as the charge degree of freedom moves (the hole), a single spin excitation is created and remains at the original site of the hole. Due to intense previous studies of these effects, we will not go any further with it in the present paper. 

For more nearest neighbors, $q \geq 3$, a series of quasiparticles peaks arises [see Figs. \ref{fig.spectral_functions}(b)--\ref{fig.spectral_functions}(d)]. In Sec. \ref{subsec.strong_interaction_limit}, we show that at strong interactions, their energies all scale with $(J_z / t)^{2/3}$. The origin of this scaling is that the system in this limit can be rephrased as a continuum model, in which the hole moves in a linear potential with strength $J_z$. While such a rephrasing in a regular lattice is approximate \cite{Bulaevskii1968,Kane1989,Grusdt2018_2}, this becomes exact in Bethe lattices as we shall unfold in Sec. \ref{subsec.strong_interaction_limit}. In the limit of weak interactions on the other hand, $J_z \gg t$, the hole becomes immobile and the $n$th eigenstate is completely localized at the $n$th depth with an energy simply given by the magnetic energy cost, $\varepsilon_n \to E_J^{(n)} = [(q-1) + (n-1) (q-2)]J_z/2$ [Eq. \eqref{eq.E_J}] for $n\geq 1$. In fact, to order $\mathcal{O}(t^2 / J_z)$, only the magnetic polaron ground state and the first string excitation are affected and attain the energies
\begin{equation}
\varepsilon^{(\pm)} = \frac{(q-1)J_z}{4}\left[1 \pm \left(1 + \frac{q}{2}\left[\frac{4t}{(q-1)J_z}\right]^2\right)\right].
\label{eq.perturbative_energies}
\end{equation}
Here, $\varepsilon_0 = \varepsilon^{(-)}$ and $\varepsilon_1 = \varepsilon^{(+)}$ are the ground and lowest string excitation energies respectively. 

The method used here to find the explicit coefficients of these eigenstates is similar to Refs. \cite{Ramsak1998,Reiter1994}, where the approximate magnetic polaron eigenstates within the SCBA \cite{Kane1989,Martinez1991,Liu1991} are found. We finally note that these many-body eigenstates can in principle be used to completely characterize the full nonequilibrium dynamics. Specifically, $\braket{\Psi_n|\Psi(\tau)} = \bra{\Psi_n} \te^{-i\hat{H}\tau}\hat{h}^\dagger_0\ket{\AF} = \te^{-i\varepsilon_n\tau}\!\bra{\Psi_n}\hat{h}^\dagger_0\ket{\AF} = \sqrt{Z_n} \cdot \te^{-i\varepsilon_n\tau}$. Therefore, the dynamics is the quantum interference of the polaron ground state and the string excitations. While this lends insight into the conceptual nature of the dynamics, it does not actually simplify the characterization of the underlying hole motion.

\subsection{Continuum limit for strong interactions} \label{subsec.strong_interaction_limit}
In this subsection, we describe in detail the continuum limit arising for $J_z/t\to 0^+$ and a number of nearest neighbors $q\geq 3$. In addition to the dominant $(J_z/t)^{2/3}$ contribution to the energy found previously \cite{Kane1989,Grusdt2018_2,Zhong1995}, we also variationally determine a term linear in $J_z$, which allows us to accurately describe the limiting behavior of the quasiparticle residue for $J_z/t\to 0^+$. Additionally, we explicitly compare the exact eigenstates found in Sec. \ref{subsec.many_body_eigenstates} to the strong-coupling limit investigated here.

We begin by transforming the equations of motion to an effective one-dimensional setting valid for the depth symmetric eigenstates found in the previous Sec. \ref{subsec.many_body_eigenstates}. To this end, we define $\psi(0) = C^{(0)}$, and $\psi(d) = (-1)^d \sqrt{q(q-1)^{d-1}} \cdot C^{(d)}$ for $d \geq 1$, similar to the approach in Ref. \cite{Grusdt2018_2}. In this manner, the wave function fulfills a one-dimensional normalization condition: $\sum_d |\psi(d)|^2 = 1$. With this in hand, the equations of motion become
\begin{align}
\!\!\!\varepsilon \psi(0) &= -\sqrt{q}t \psi(1), \nn \\
\!\!\!\varepsilon \psi(1) &= E_J^{(1)} \psi(1) - \sqrt{q}t \psi(0) - \sqrt{q-1}t \psi(2), \nn \\
\!\!\!\varepsilon \psi(d) &= E_J^{(d)} \psi(d) - \sqrt{q-1}t \left[\psi(d-1) + \psi(d+1)\right],
\label{eq.equations_of_motion_1}
\end{align}
where the lower equation is valid for $d\geq 2$. We may now formulate the continuum model by taking this lower equation and rephrase it as a continuous equation in the limit of $J_z \ll t$. This is possible, because the wave function spreads out over an ever increasing number of lattice sites in this limit, as we shall explicitly see in Sec. \ref{sec.results}. Here, it is first beneficial to rewrite this equation according to
\begin{align}
\tilde{\varepsilon}\psi(d) =& \, \frac{(q-2)J_z}{2\sqrt{q-1}t } d \cdot \psi(d) \nn \\
& - \left[\psi(d-1) - 2\psi(d) + \psi(d+1)\right],
\label{eq.equations_of_motion_2}
\end{align}
using Eq. \eqref{eq.E_J}, and defining $\sqrt{q-1}t \cdot \tilde{\varepsilon} = \left[\varepsilon + 2\sqrt{q-1}t \right]$. Next, we let $d = x / \lambda$ and introduce a rescaled wave function $\phi(x) = \psi(x / \lambda) / \sqrt{\lambda}$. Inserting this in Eq. \eqref{eq.equations_of_motion_2} and dividing out $\lambda^2$, we get
\begin{align}
\frac{\tilde{\varepsilon}}{\lambda^2}\phi(x) =& \, \frac{(q-2)J_z}{2\sqrt{q-1}t} \frac{x}{\lambda^3} \phi(x) \nn \\ &- \frac{\phi(x-\lambda) - 2\phi(x) + \phi(x+\lambda)}{\lambda^2}.
\label{eq.equations_of_motion_3}
\end{align}
To eliminate $J_z / t$, we set $\lambda = [(q-2)J_z/(2\sqrt{q-1}t)]^{1/3}$. For $J_z / t \to 0^+$, the fraction in the second line of Eq. \eqref{eq.equations_of_motion_3} approaches the second order derivative. Setting $a = \tilde{\varepsilon} / \lambda^2$, we, hereby, obtain
\begin{equation}
a\phi(x) = x \cdot \phi(x) - \frac{{\rm d}^2\phi}{{\rm d}x^2},
\label{eq.equations_of_motion_4}
\end{equation}
which is accurate up to order $\lambda^2 \propto (J_z / t)^{2/3}$. The wave function $\phi(x)$ fulfills the continuous normalization condition: $\int_{0}^\infty\!{\rm d}x |\phi(x)|^2 = 1$. Letting $y = x - a$, we get the Airy equation
\begin{equation}
0 = y f(y) - \frac{{\rm d}^2f}{{\rm d}y^2},
\label{eq.airy_equation}
\end{equation}
for $y \geq - a$, where $f(y) = \phi(y + a)$. The solutions to this equation can thus be written as a superposition of the Airy functions: $f(y) = A \cdot {\rm Ai}(y) + B \cdot {\rm Bi}(y)$. Since ${\rm Bi}(y)$ blows up for $y \to \infty$, $B = 0$. Furthermore, the continuous one-dimensional description, which is exact for $J_z / t \to 0^+$, entails that $\phi(x)$ must vanish for $x < 0$, and therefore also at $x = 0$. In turn, $-a$ must be a zero of the Airy function ${\rm Ai}(y)$. In this way, we obtain the sought set of eigenstates in the strongly interacting limit
\begin{align}
\psi_n(d) &= \sqrt{\lambda}\cdot \phi_n\left(\lambda d\right) = \sqrt{\lambda} \cdot  A_n {\rm Ai}\left(\lambda d - a_n\right).
\label{eq.eigenstates_strong}
\end{align}

%%%%%%%%%%%%%%%%%%%%%%%%%%%%%%%%%%%%%%%%%%%%%%%%%%%%%%%%%%%%%%%%%%% 
\begin{figure}[t!]
\begin{center}
\includegraphics[width=0.95\columnwidth]{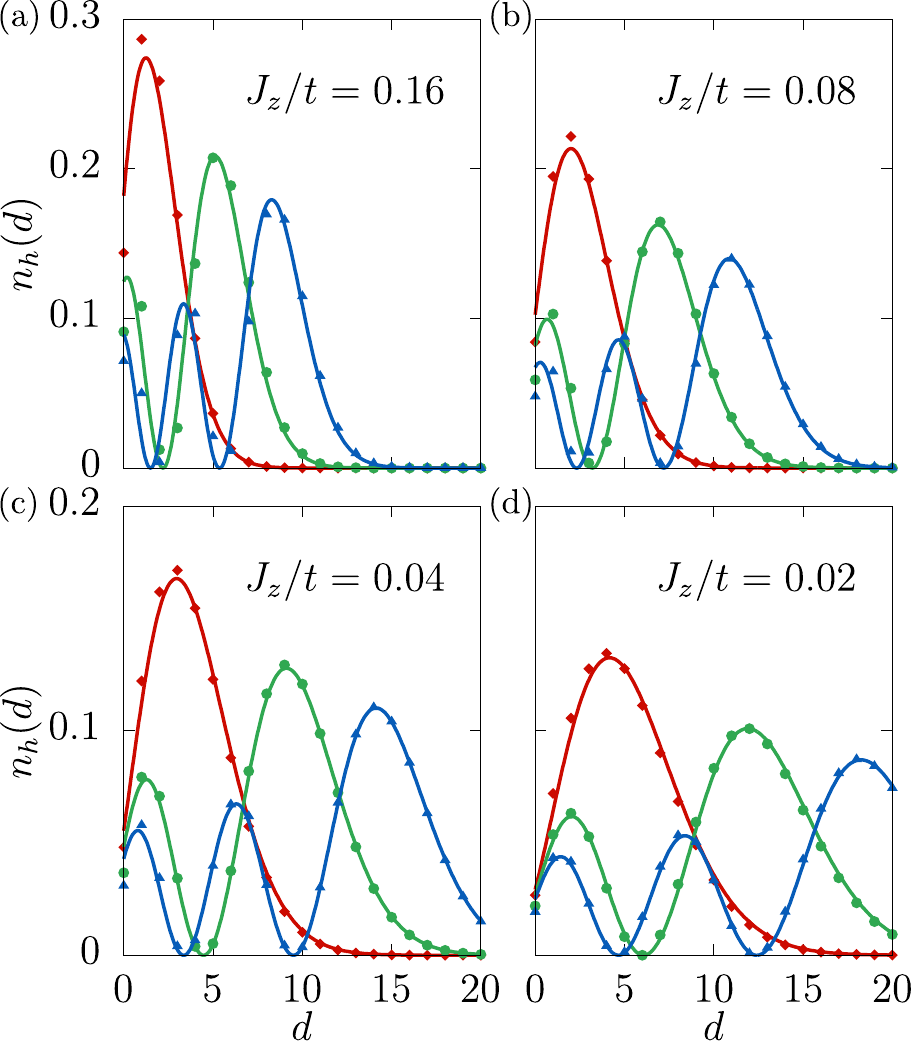}
\end{center}\vspace{-0.5cm}
\caption{\textbf{Eigenstates for strong interactions.} The hole density is plotted as function of depth for four indicated interaction strengths [(a)--(d)] and $q = 4$ nearest neighbors. We compare the numerically calculated eigenstates from Eq. \eqref{eq.C_d_n_general} to the strongly interacting limit, Eq. \eqref{eq.eigenstates_strong} with $d \to d - d_0$, for the three lowest string excitation states in red (ground state), green (first excited state), and blue (second excited state).}
\vspace{-0.25cm}
\label{fig.eigenstates_strong} 
\end{figure} 
%%%%%%%%%%%%%%%%%%%%%%%%%%%%%%%%%%%%%%%%%%%%%%%%%%%%%%%%%%%%%%%%%%% 

The $n$th eigenstate is, thus, defined by the $n$th order zero of the Airy function $-a_n$. Also, $A_n^{-2} = \int_0^\infty {\rm d}x\, |{\rm Ai}\left(x - a_n\right)|^2$ ensures normalized eigenstates. For any nonzero value of $J_z / t$, the eigenstates remain nonzero at the origin, $d = 0$. This turns out to yield a correction linear in $J_z / t$. To accommodate for this, we let $d \to d - d_0$ in Eq. \eqref{eq.eigenstates_strong} and use $d_0$ as a variational parameter. A rather lengthy calculation (see Appendix \ref{app.strong_eigenstates_d0}) yields the variational energy for the $n$th eigenstate
\begin{align}
\!\varepsilon_n^{\rm var} &= - 2\sqrt{q-1}t\left(1 - \frac{a_n}{2} \lambda^2\right) + c_q(d_0) J_z,
\label{eq.eigenenergies_order_1}
\end{align}
with $c_q(d_0) = (q-2)[d_0 + (q-2)^{-1} + 2(\sqrt{q(q-1)^{-1}} - 1) d_0(1-d_0) + (d_0+1/2)^2]$. The two first terms yield the exact asymptotic behavior \cite{Zhong1995} with the dominant $\lambda^2 \propto (J_z / t)^{2/3}$ scaling. The term proportional to $J_z$ is variationally determined, yielding a value of $d_0 = 1 / (2 - 3\sqrt{q(q-1)^{-1}})$. The resulting energies of the three lowest energies are plotted as colored lines in Fig. \ref{fig.spectral_functions} for $q = 3$, $4$ and $5$. The corresponding eigenstates for $q = 4$ are plotted in Fig. \ref{fig.eigenstates_strong}, showing the approach to the strongly interacting limit. Whereas the behavior at large depths is already captured very well at $J_z / t = 0.16$, the convergence of the full wave function requires very strong interactions of around $J_z / t = 0.02$.

%%%%%%%%%%%%%%%%%%%%%%%%%%%%%%%%%%%%%%%%%%%%%%%%%%%%%%%%%%%%%%%%%%% 
\begin{figure}[t!]
\begin{center}
\includegraphics[width=0.85\columnwidth]{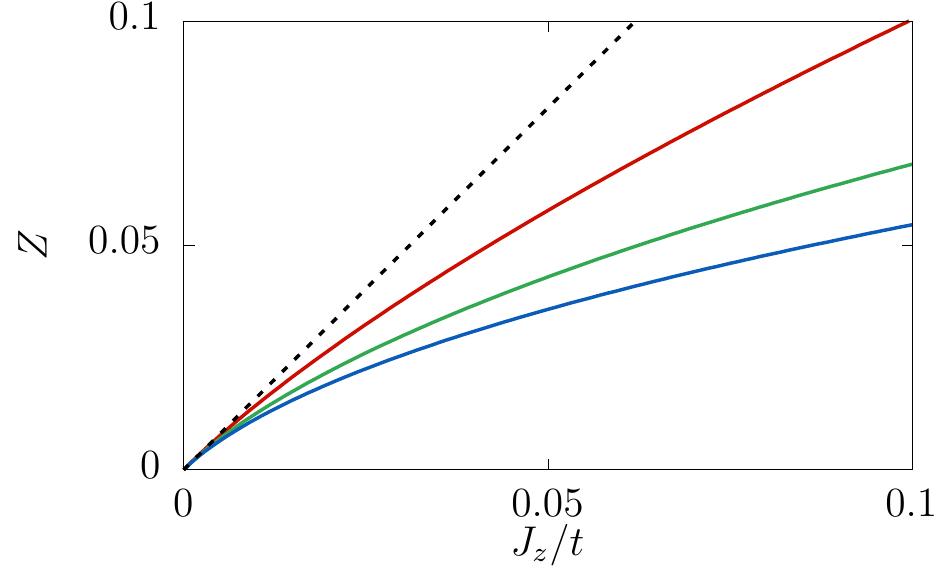}
\end{center}\vspace{-0.5cm}
\caption{\textbf{Residue for strong interactions.} The residue is plotted for three lowest energy states as a function of interaction strength: the ground state in red, the first excited state in green, and the second excitated state in blue. This is compared to the linear behavior in Eq. \eqref{eq.residue_asymptote} (black dashed), confirming the universal linear trend at strong enough interactions.}
\label{fig.residue_strong} 
\vspace{-0.25cm}
\end{figure} 
%%%%%%%%%%%%%%%%%%%%%%%%%%%%%%%%%%%%%%%%%%%%%%%%%%%%%%%%%%%%%%%%%%% 

The presence of a nonzero $d_0$ allows us to to calculate the residues, $Z_n$, in this asymptotic strong coupling limit. This yields
\begin{align}
Z_n = |\psi_n(d = 0)|^2 \simeq \lambda \cdot \left[A_n {\rm Ai}\left(-\lambda d_0 - a_n\right)\right]^2
\label{eq.residue_approximate}
\end{align}
This can be greatly simplified by expanding the Airy function. In fact, ${\rm Ai}\left(-\lambda d_0 - a_n\right) \to (-\lambda d_0) {\rm Ai}'(-a_n)$, for $\lambda \propto (J_z / t)^{1/3} \to 0$. Furthermore, using the integral relation $\int_0^\infty {\rm d}x [{\rm Ai}(x - a_n)]^2 = a_n [{\rm Ai}(-a_n)]^2 + [{\rm Ai}'(-a_n)]^2 = [{\rm Ai}'(-a_n)]^2$, and that $A_n^{-2} = \int_0^\infty {\rm d}x [{\rm Ai}(x - a_n)]^2 = [{\rm Ai}'(-a_n)]^2$ for $(J_z / t)^{1/3} \to 0$, we get the very simple asymptotic behavior of the residues
\begin{align}
Z_n &\to \lambda \cdot \left[A_n (-\lambda d_0) {\rm Ai}'(-a_n)\right]^2 = \lambda^3 d_0^2 \nn \\
&= \frac{q-2}{2\sqrt{q-1}}\frac{1}{\left(2 - 3 \sqrt{\frac{q}{q-1}}\right)^2} \cdot \frac{J_z}{t},
\label{eq.residue_asymptote}
\end{align}
linear in $J_z / t$ as argued previously \cite{Kane1989}. Additionally, we find it to be independent of the eigenstate $n$, essentially because the energy shifts due to a nonzero $d_0$, $\varepsilon_n^{\rm var} - \varepsilon_n$ is independent of $n$. Consequently, this variational approach strongly suggests that the residues of all string excitation states approach this universal value, which is confirmed by our numerical calcuations in Fig. \ref{fig.residue_strong} for $q = 4$ nearest neighbors. Consequently, the quasiparticles remain well-defined for any $J_z/t>0$. Note that the deviations between the full result and the continuum limit at first glance seem larger for the residues in Fig. \ref{fig.residue_strong} than for the overall eigenstates in Fig. \ref{fig.eigenstates_strong}. However, a closer inspection of Fig. \ref{fig.eigenstates_strong} reveals that the short-range part of the eigenstates, and in particular the residue $Z_n = |\bra{\Psi_n}\hat{h}^\dagger_0\ket{\AF}|^2$, only approach the continuum limit at very low values of $J_z / t$, as we might expect from comparing to a continuum limit. In more detail, while the long-range part of the eigenstates are determined by the $(J_z / t)^{2/3}$ term in the energy [see Eq. \eqref{eq.eigenenergies_order_1}], the finite value of the residue arises due to the linear term, and is, therefore, more prone to higher-order corrections.  

\section{Nonequilibrium hole and spin dynamics} \label{sec.results}
In this section, we describe the main results for the nonequilibrium dynamics. In Section \ref{subsec.hole_dynamics}, we investigate the hole dynamics, whereas the nonequilibrium spin-spin correlation dynamics is studied in Sec. \ref{subsec.spin_dynamics}. 

\subsection{Hole dynamics} \label{subsec.hole_dynamics}
\begin{figure}[t!]
\begin{center}
\includegraphics[width=1.0\columnwidth]{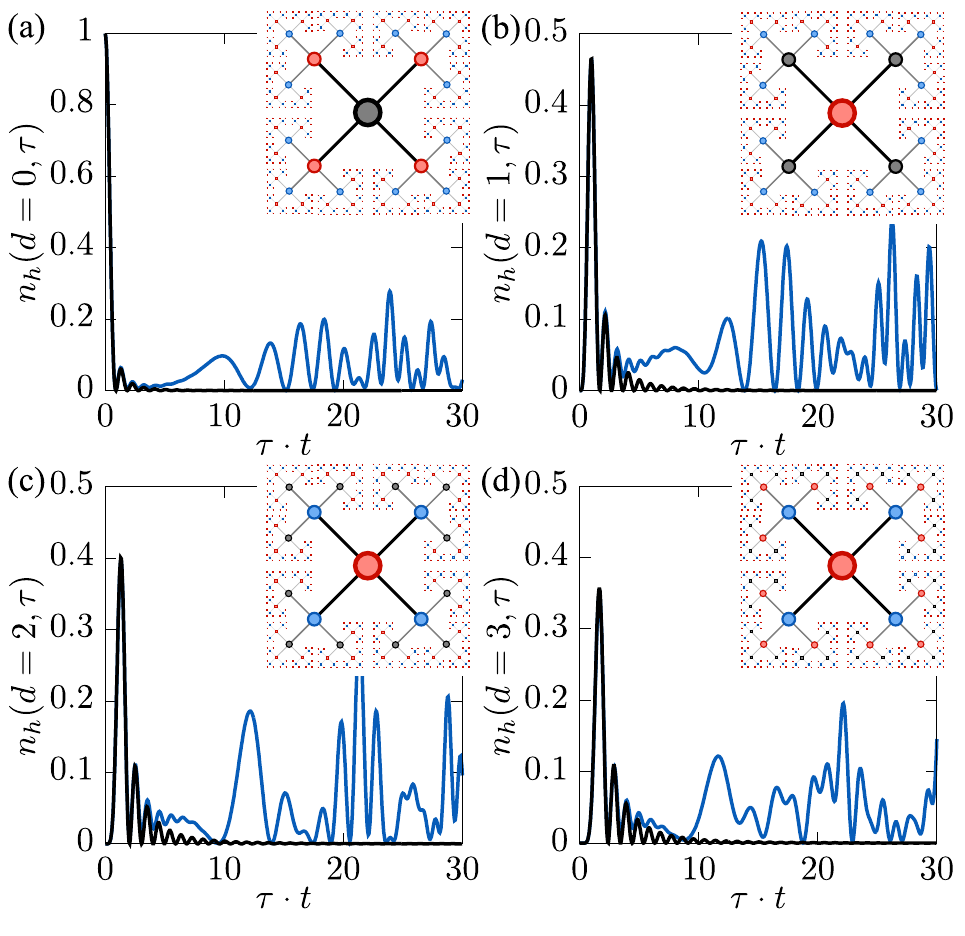}
\end{center}\vspace{-0.5cm}
\caption{\textbf{Local densities}. We plot the local density of holes in the vicinity of its original position, $d = 0$, for 4 indicated depths (insets) in the case of $q = 4$ nearest neighbors, and a spin-spin coupling $J_z = 0.2t$ (blue lines). This is further compared to the limit of infinite interactions, $J / t \to 0^+$, where the hole moves like a free particle (black lines). While the initial stage shows a clear dampening of regular oscillations, associated with a free quantum walk, for any nonzero $J_z/t$ large aperiodic oscillations start to kick at later times, in the present case of $J_z / t = 0.2$ around $\tau = 10 / t$.}
\label{fig.local_densities} 
\vspace{-0.25cm}
\end{figure}

The density of holes at certain depth $d$ is readily computed from the coefficients of the many-body wave function as
\begin{align}
n_h(d = 0, \tau) &= |C^{(0)}(\tau)|^2, \nn \\
n_h(d, \tau) &= q(q-1)^d \cdot |C^{(d)}(\tau)|^2, \;\; d\geq 1.
\label{eq.n_h_d_and_C_d}
\end{align}
Here, $C^{(d)}(\tau)$ is calculated from Eq. \eqref{eq.C_d_and_R_d}. The results for $q = 4$ and $J_z = 0.2t$ are shown in Fig. \ref{fig.local_densities} in the vicinity of the initial position of the hole. Since $C^{(d)}(\tau)$ is the Fourier transform of $2\Re[R^{(d)}(\omega)]$, given in Eq. \eqref{eq.R_d_general}, the appearance of heavy oscillations in these local densities can be understood from the presence of a plethora of spectral peaks, Fig. \ref{fig.spectral_functions}(d), corresponding to the string excitations found in Sec. \ref{subsec.many_body_eigenstates}. Furthermore, the presence of a hole at depth $d$ is always accompanied by $d$ overturned spins, see Eqs. \eqref{eq.exact_many_body_wave_function} and \eqref{eq.correlation_functions}, and, therefore, entails a spin string of length $d$. As a result, the hole density distribution is equal to the probability distribution for the so-called string length, which has been investigated previously in a two-dimensional square lattice both theoretically \cite{Grusdt2018_2,Bieniasz2019} and experimentally \cite{Chiu2019}.  

%%%%%%%%%%%%%%%%%%%%%%%%%%%%%%%%%%%%%%%%%%%%%%%%%%%%%%%%%%%%%%%%%%% 
%%%%%%%%%%%%%%%%%%%%%%%%%%%%%%%%%%%%%%%%%%%%%%%%%%%%%%%%%%%%%%%%%%% 

%%%%%%%%%%%%%%%%%%%%%%%%%%%%%%%%%%%%%%%%%%%%%%%%%%%%%%%%%%%%%%%%%%% 
\begin{figure}[t!]
\begin{center}
\includegraphics[width=1.0\columnwidth]{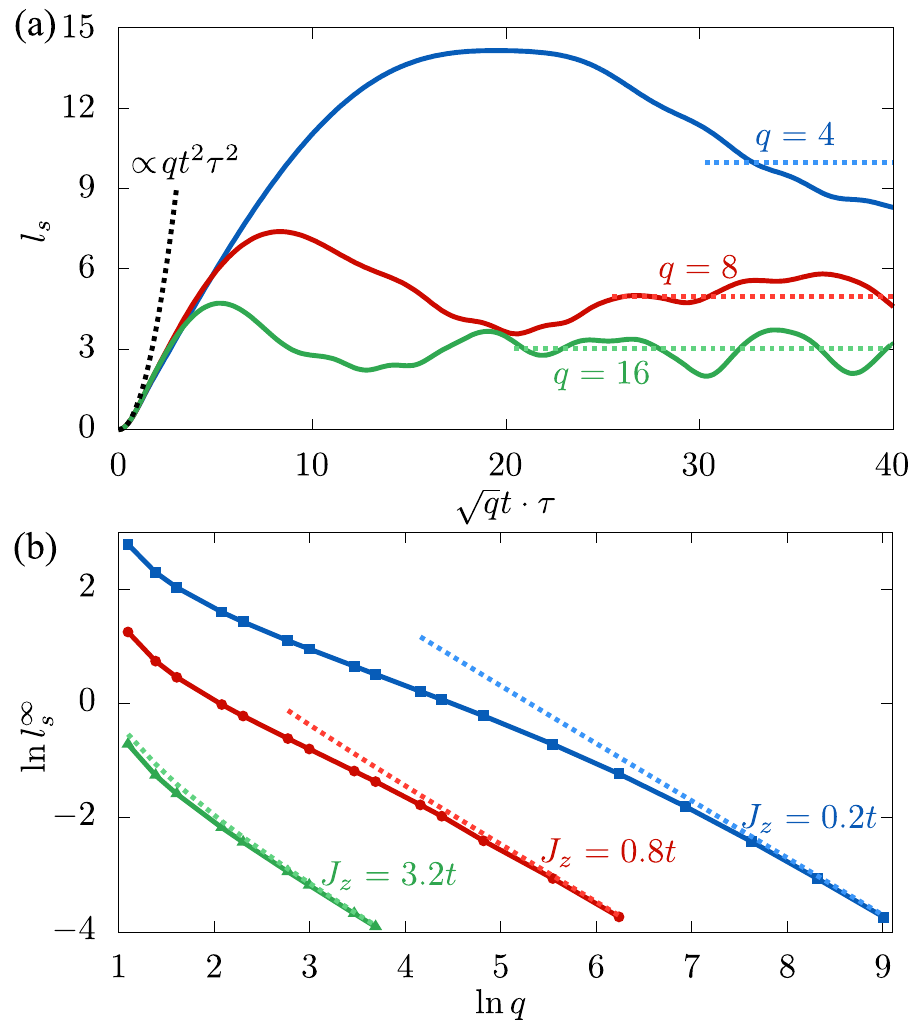}
\end{center}\vspace{-0.5cm}
\caption{\textbf{String length vs. coordination number}. (a) The string length $l_s(\tau) = \sum_d d \cdot n_h(d, \tau)$ as a function of time for three different coordination numbers, $q = 4, 8, 16$. For $\tau \ll 1 / \sqrt{q}t$, all curves collapse to an initial ballistic behavior of the hole, corresponding to $l_s = qt^2\cdot \tau^2$. At long times, the string length has heavy oscillations around a well-defined mean value, $l_s^\infty$, shown in dashed lines. This mean value is plotted in (b) as a function of the coordination number $q$ for three different values of $J_z / t$, and compared to the weak-coupling result (dashed lines).}
\label{fig.string_length_q} 
\vspace{-0.25cm}
\end{figure} 
%%%%%%%%%%%%%%%%%%%%%%%%%%%%%%%%%%%%%%%%%%%%%%%%%%%%%%%%%%%%%%%%%%%

To better understand this complex many-body scenario, we calculate the average depth of the hole,
\begin{equation}
l_s(\tau) = \sum_d d\cdot n_h(d,\tau).
\label{eq.string_length}
\end{equation}
We denote it $l_s$, as it also gives the average \emph{length of overturned spins}, or simply the string length, at time $\tau$. This is plotted in Fig. \ref{fig.string_length_q}(a) as a function of time for three indicated coordination numbers. Solving the equations of motion at short times, reveals that the initial dynamics is a free quantum walk independent of the inverse interaction strength $J_z / t$, with 
\begin{equation}
l_s = qt^2 \cdot \tau^2 + \mathcal{O}[(t\cdot\tau)^4]. 
\label{eq.l_s_short_times}
\end{equation}
It is actually fairly easy to show \cite{Nielsen2022} that the initial motion of an initially localized hole within the $t$-$J$ model has to be that of a free quantum walk, even in the presence of anisotropic spin-couplings. 

%%%%%%%%%%%%%%%%%%%%%%%%%%%%%%%%%%%%%%%%%%%%%%%%%%%%%%%%%%%%%%%%%%% 
\begin{figure}[t!]
\begin{center}
\includegraphics[width=1.0\columnwidth]{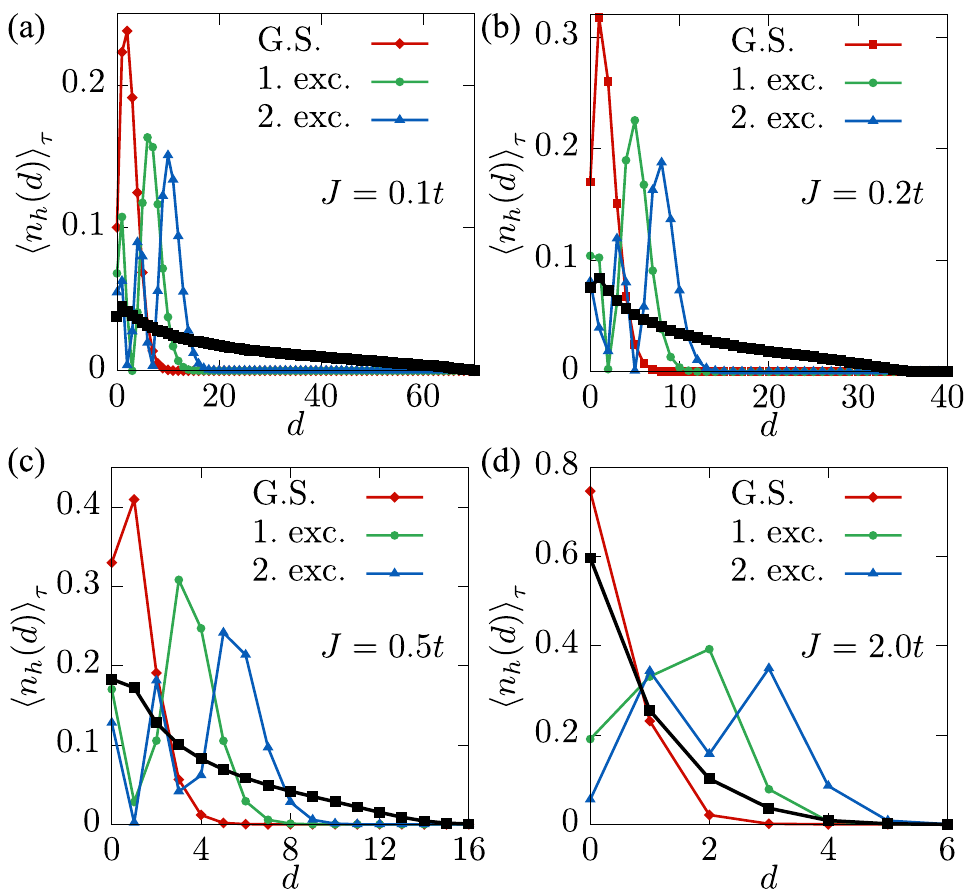}
\end{center}\vspace{-0.5cm}
\caption{\textbf{Time-averaged hole distribution}. In black squares is shown the time-averaged hole density distribution $\braket{n_h(d)}_\tau$ [Eq. \eqref{eq.average_hole_density}] compared to the hole distribution for the first 3 eigenstates (colored markers) as a function of the depth $d$ and for 4 indicated interaction strengths. The coordination number is set to $q = 4$.}
\label{fig.ave_density_distributions} 
\vspace{-0.25cm}
\end{figure} 
%%%%%%%%%%%%%%%%%%%%%%%%%%%%%%%%%%%%%%%%%%%%%%%%%%%%%%%%%%%%%%%%%%%

\begin{figure}[t!]
\begin{center}
\includegraphics[width=1.0\columnwidth]{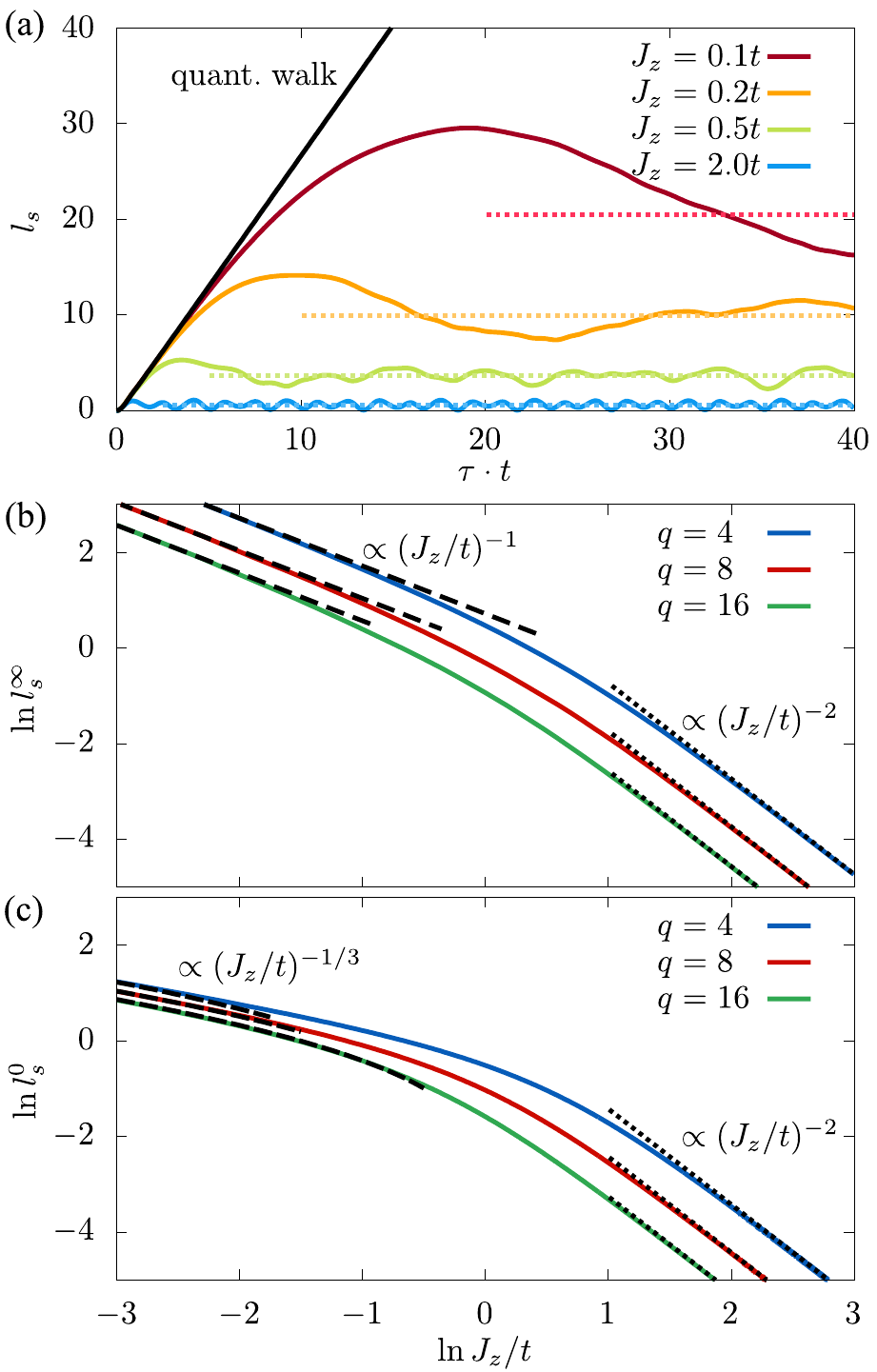}
\end{center}\vspace{-0.5cm}
\caption{\textbf{String length vs. interaction strength}. (a) The string length is plotted as a function of time for $q = 4$ for four indicated inverse interaction strengths, $J_z / t$, with long time averages, $l_s^{\infty}$, shown as dashed lines. In the limit of infinitely strong interactions, $J_z / t = 0^+$, the hole motion is a free quantum walk [black solid line]. (b) The long time averages, $l_s^{\infty}$, is plotted as a function of interaction strength and further compared to the string length in the magnetic polaron ground state $l_s^{0}$ (c) for three indicated values of the coordination number $q$. At weak coupling, $l_s^{\infty} = 2 \cdot l_s^0 \propto (J_z / t)^{-2}$ (black short-dashed lines). At strong coupling, they feature different power-law scalings (black long-dashed lines) with $l_s^{\infty} \propto (J_z / t)^{-1}$ and $l_s^0 \propto (J_z / t)^{-1/3}$.}
\label{fig.string_length_J} 
\vspace{-0.25cm}
\end{figure}

At long times, our results reveal that the string length undergoes heavy oscillations around a well-defined mean value, 
\begin{equation}
l_s^\infty = \sum_d d\cdot \braket{n_h(d)}_\tau,
\label{eq.string_length_long_time_average}
\end{equation}
where 
\begin{equation}
\braket{n_h(d)}_\tau = \lim_{T\to\infty}\frac{1}{T} \int_0^T {\rm d}\tau \; n_h(\bd, \tau). 
\label{eq.average_hole_density}
\end{equation}
Physically, this finite asymptote reflects that the hole remains bound to its initial position. The time-averaged hole distribution in Eq. \eqref{eq.average_hole_density} is shown in Fig. \ref{fig.ave_density_distributions} for a set of indicated inverse interaction strengths $J_z / t$, and compared to the hole distribution for the polaron ground state and the two lowest string states [see Sec. \ref{subsec.many_body_eigenstates}]. It is evident that for strong coupling, $J_z \ll t$, the dynamical wave function is significantly more spread out than its equilibrium counterparts. This is a natural consequence of the fact that the average energy of the quenched system $\bra{\Psi(\tau)} H \ket{\Psi(\tau)} = 0$ (relative to $E_J^{(0)} = qJ_z / 2 + E_0(q,d)$) is much larger than the ground state energy $\sim-2\sqrt{q-1}t$. Additionally, the shape of the time-averaged distribution changes quite dramatically with decreasing $J_z / t$. Indeed, below $J_z = 0.4t$ the hole is no longer found with the highest probability at its original site, but rather one of its nearest neighbors. In Fig. \ref{fig.string_length_q}(b), we compare the asymptotic string length $l_s^\infty$ to the weak coupling result, 
\begin{equation}
l_s^\infty \to \frac{8q}{(q-1)^2}\cdot \left(\frac{J_z}{t}\right)^{-2},
\label{eq.l_s_weak_coupling}
\end{equation} 
valid for $J_z / t \gg 1$. To further investigate the dependency on the interaction strength, we plot the string length dynamics in Fig. \ref{fig.string_length_J}(a) for several indicated values of $J_z / t$. As can be expected, the hole travels further into the Bethe lattice for decreasing values of $J_z / t$. The motion is generally aperiodic, due to the irregular spacing of the energy levels $\varepsilon_n$ evident in Fig. \ref{fig.spectral_functions}. The only exception is in the limit of very weak interactions, $J_z \gg t$, in which the hole is restricted to hop back and forth between depths $d = 0$ and $d = 1$, resulting in periodic motion with an angular frequency given by the energy difference between the polaron ground state and the lowest string excitation, 
\begin{equation}
\varepsilon_1 - \varepsilon_0 \to (q-1)\frac{J_z}{2}\left(1 + \frac{q}{2}\left[\frac{4t}{(q-1)J_z}\right]^2\right), 
\label{eq.perturbative_angular_frequency}
\end{equation}
approaching the magnetic energy cost of going to depth $d = 1$, $E_J^{(1)} = (q-1)J_z / 2$, as $J_z \gg t$. In the opposite extreme of $J_z / t = 0^+$, the dynamics is characterized by a free quantum walk of the hole as anticipated by Eqs. \eqref{eq.G_1_J_0_limit} and \eqref{eq.J_0_limit}. In Figures \ref{fig.string_length_J}(b) and \ref{fig.string_length_J}(c), we further characterize the full dependency on the inverse interaction strength, $J_z / t$. Whereas the two string lengths are simply proportional in the weak coupling limit with $l_s^\infty = 2\cdot l_s^0$, they feature remarkably different scaling behaviors for strong coupling. In fact, for any number of nearest neighbors, $q \geq 3$, our power-law fits at strong interactions, $J_z / t \ll 1$, reveal that $l_s^\infty = f^\infty(q) \cdot (J_z / t)^{-1}$. On the contrary, the scaling law for the eigenstates are dramatically different, as we may derive explicitly from the strong coupling states derived in Sec. \ref{subsec.strong_interaction_limit}. 

%%%%%%%%%%%%%%%%%%%%%%%%%%%%%%%%%%%%%%%%%%%%%%%%%%%%%%%%%%%%%%%%%%% 
%%%%%%%%%%%%%%%%%%%%%%%%%%%%%%%%%%%%%%%%%%%%%%%%%%%%%%%%%%%%%%%%%%%  

Explicitly, we can use that the $n$th eigenstate is asymptotically given by $\psi_n(d) = \sqrt{\lambda}\phi_n(\lambda (d-d_0)) = \sqrt{\lambda}A_n \cdot {\rm Ai}(\lambda(d-d_0)-a_n)$. Here, we also include the effect of a nonzero shift $d_0$. We then get 
\begin{align}
l_s^n &= \sum_d d \cdot |\psi_n(d)|^2 = \lambda \sum_d d \cdot |\phi_n(\lambda (d-d_0))|^2 \nn \\ &= \sum_{x\geq -\!\lambda d_0} \!\!\!\left(\frac{x}{\lambda} + d_0 \right) |\phi_n(x)|^2 \Delta x. \nn
\end{align}
Here, we use $x = \lambda(d-d_0)$, whereby $\Delta x = \lambda$. The term proportional to $d_0$ is simply the normalization of the wave function, and so just yields $d_0$. The remaining terms can, in the limit of strong interactions $\lambda \propto (J_z/t)^{1/3} \to 0^+$, be rephrased as an integral. This yields
\begin{align}
l_s^n &\to d_0 + \frac{1}{\lambda}\int_0^\infty {\rm d}x \, x |\phi_n(x)|^2 \nn \\
&= d_0 + \frac{1}{\lambda}\int_0^\infty {\rm d}x \, x A_n^2 [{\rm Ai}(x-a_n)]^2 \nn \\
&= d_0 + \frac{1}{\lambda} \cdot \frac{2a_n}{3} A_n^2 [{\rm Ai}'(-a_n)]^2 = d_0 + \frac{2a_n}{3\lambda}.
\label{eq.l_s_n}
\end{align}
In the last line, we first use an integral relation for the Airy functions: $\int_0^\infty {\rm d}x \, x A_n^2 [{\rm Ai}(x-x_0)]^2 = (2x_0 [{\rm Ai}(-x_0)]^2 - {\rm Ai}(-x_0){\rm Ai}'(-x_0) + 2 x_0 [{\rm Ai}'(-x_0)]^2)/3$, and that $-x_0 = -a_n$ is a zero of the Airy function. Finally, we use that the normalization constant is given by $A_n^{-2} = [{\rm Ai}'(-a_n)]^2$. This expression, hereby, yields a dominant $\lambda^{-1} \propto (J_z/t)^{-1/3}$ scaling of the string length for all eigenstates. Additionally, the increase in string length for eigenstates with higher energy is simply linearly related to the increase in the zeros of the Airy function $-a_n$. In Fig. \ref{fig.string_length_J}, Eq. \eqref{eq.l_s_n} is compared to the numerically obtained string length for the ground state for three different values of the number of nearest neighbors, showing excellent agreement at strong coupling. 

We note that to get converging results for the thermodynamic limit in the case of $q = 4$ and very strong interactions of $J_z = 0.05t$, we need to go to a total depth of at least $d_{\rm tot} = 200$. In this case, the Bethe lattice consists of $N(q = 4, d_{\rm tot} = 200) \simeq 10^{95}$ sites [Eq. \eqref{eq.number_of_sites}]. This far exceeds the total number of atoms in the observable universe \cite{LifeScience_atoms_in_obs_universe}, and exemplifies the enormity of the simplification achieved when reducing the description of an exponential number of sites in the Bethe lattice with just a linear number of coefficient, $C^{(d)}$. 

\subsection{Spin dynamics} \label{subsec.spin_dynamics}
In the present subsection, we investigate the dynamics of the spin-spin correlation function
\begin{equation}
C_S(d, \tau) = 4\bra{\Psi(\tau)}\hat{S}^{(z)}_{0}\hat{S}^{(z)}_{{\bf j}_d}\ket{\Psi(\tau)}.
\label{eq.spin_correlator}
\end{equation}
This describes the tendency of the spin at the origin, $d = 0$, to align ($C_S > 0$) or antialign ($C_S < 0$) with a spin at depth $d$. Note that the depth symmetry of the dynamics entails that $C_S$ only depends on the depth $d$ of the second spin. The advent of quantum simulation platforms enables the study of such quantities, as has been seen in two-dimensional square lattices both in \cite{Koepsell2019} and out of equilibrium \cite{Ji2021}. On the other hand, the actual computation of these correlators often present an astonishing theoretical feat. In the presently studied Bethe structures, however, the full knowledge of the many-body wave function enables the precise and efficient investigation of the spin-spin correlator in Eq. \eqref{eq.spin_correlator}. 

To see this more concretely, we link $C_S$ to the coefficients of the many-body wave function. In the absence of a hole, the system is a perfect antiferromagnetic state, resulting in $C_S^{(0)}(d) = 4\bra{\AF}\hat{S}^{(z)}_{0}\hat{S}^{(z)}_{{\bf j}_d}\ket{\AF} = (-1)^{d}$. This overall sign expresses the perfectly staggered antiferromagnetism. In the presence of a hole, we now link $C_S(d, \tau)$ to the hole density. Consider, then, first the case where the hole is located between depths $0$ and $d$, $0 < d_h < d$. In this case, the $z$ component of the spin at $d = 0$ is just $+ 1 / 2$, while the $z$ component of the spin at depth $d$ is $(-1)^{d-1} / 2$. In turn, we get a contribution of $(-1)^{d-1} \cdot P(0 \!<\! d_h \!<\! d, \tau)$ to $C_S(d, \tau) $. Here, $P(0\!<\!d_h\!<\!d,\tau) = \sum_{0<d_h<d} n_h(d_h, \tau)$ is the probability to find the hole between depths $0$ and $d$ at time $\tau$. 

Next, if the hole has passed depth $d$, the above correlation flips sign if the hole has passed the specific site ${\bf j}_d$. If not, then the correlation does not flip. The relative probability to have passed ${\bf j}_d$ is just $1 / q(q-1)^{d-1}$. If the hole passes ${\bf j}_d$, there is, thus, a contribution of $(-1)^{d} \cdot P(d_h \!>\! d, \tau) / q(q-1)^{d-1}$. If it does \emph{not} pass ${\bf j}_d$, it contributes with $(-1)^{d-1} \cdot P(d_h \!>\! d, \tau) (1 - 1/ q(q-1)^{d-1})$. Here, $P(d_h \!>\! d, \tau) = \sum_{d_h > d}n_h(d_h, \tau)$ is the probability for the hole to have passed depth $d$. 

Finally, if the hole is at the specific depth $d$, the correlator $C_S(d, \tau)$ vanishes if the hole is at site ${\bf j}_d$. The contribution from this scenario is, therefore, only $(-1)^{d-1} \cdot P(d_h = d, \tau) (1 - 1/ q(q-1)^{d-1})$, coming from the instance where the hole \emph{is not} at ${\bf j}_d$. In total then, the nonequilibrium spin-spin correlator in Eq. \eqref{eq.spin_correlator} is
\begin{align}
&C_S(d, \tau) = (-1)^{d-1} \Bigg[P(0 < d_h < d, \tau) \nn \\
&\phantom{C_S(d, \tau) =}+ P(d_h > d, \tau)\left(1 - \frac{2}{q(q-1)^{d-1}}\right) \nn\\
&\phantom{C_S(d, \tau) =}+ P(d_h = d, \tau)\left(1 - \frac{1}{q(q-1)^{d-1}}\right) \Bigg] \nn \\
&\!= \!(-1)^{d-1} \Bigg[1 \!-\! n_h(0, \tau) \!-\! \frac{n_h(d, \tau) + 2\sum_{d_h\! > d} n_h(d_h, \tau)}{q(q-1)^{d-1}} \Bigg].
\label{eq.spin_correlator_result}
\end{align}
This expression is valid for any $d\geq 1$. For $d = 0$, we simply have $C_S(d = 0, \tau) = 1 - n_h(d = 0,\tau)$. In Eq. \eqref{eq.spin_correlator_result}, we use that the total probability of finding the hole away from the origin is $1$ minus the hole density at $d = 0$: $P(0 \!<\! d_h \!<\! d, \tau) + P(d_h \!=\! d, \tau) + P(d_h \!>\! d, \tau) = P(d_h \!>\! 0, \tau) = 1 - n_h(d_h = 0, \tau)$. At $\tau = 0$, it follows that $C_S(d, \tau = 0) = 0$, which also has to be the case physically, because there is no spin at $d = 0$ initially. At later times, as $n_h(0, \tau)$ diminishes, the spin-spin correlations can be strongly affected in the vicinity of $d = 0$. If e.g. the hole is entirely located at $d = 1$, $C_S(d = 1, \tau) = 1 - 1 / q > 0$. This has the opposite sign of the spin correlations in the absence of holes, and is simply a result of removing the original spin-$\downarrow$ fermion at $d = 0$, and letting the resulting hole travel to $d = 1$ [see Fig. \ref{fig.bethe_lattices_broken_bonds}]. 

%%%%%%%%%%%%%%%%%%%%%%%%%%%%%%%%%%%%%%%%%%%%%%%%%%%%%%%%%%%%%%%%%%% 
\begin{figure}[t!]
\begin{center}
\includegraphics[width=1.0\columnwidth]{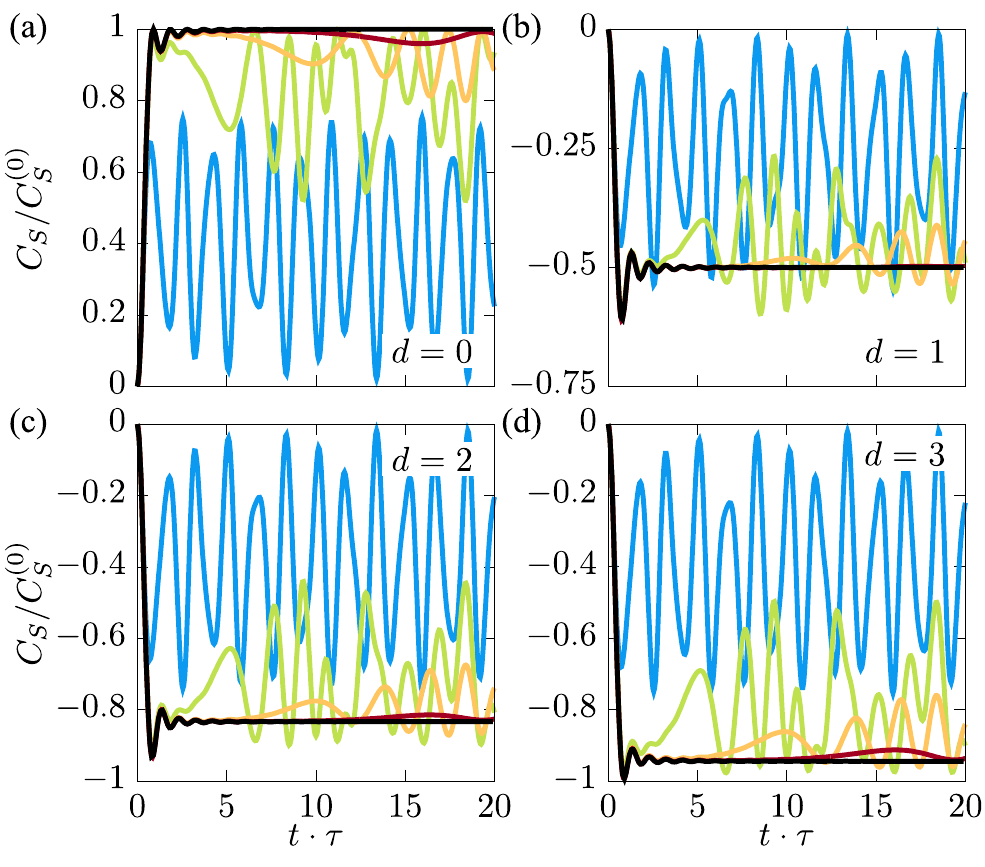}
\end{center}\vspace{-0.5cm}
\caption{\textbf{Spin dynamics}. The time-dependent spin-spin correlation function $C_S(d,\tau) = 4\bra{\Psi(\tau)}\hat{S}^{(z)}_0\hat{S}^{(z)}_{{\bf j}_d}\ket{\Psi(\tau)}$ [Eqs. \eqref{eq.spin_correlator} and \eqref{eq.spin_correlator_result}] relative to the spin correlation in the absence of a hole $C_S^{(0)}(d) = (-1)^d$ for indicated depths $d$ [(a)--(d)]. This is shown in the case of weak ($J = 2t$, blue lines), intermediate ($J = 0.5t$, green lines), strong ($J = 0.2t$, orange lines), and very strong interactions ($J = 0.1t$, red lines), as well as in the quantum walk limit of $J / t \to 0^+$ (black lines). Note that for all $d\geq 1$, the spin correlation has flipped sign with respect to the value in the absence of holes.}
\label{fig.spin_dynamics}
\vspace{-0.25cm} 
\end{figure} 
%%%%%%%%%%%%%%%%%%%%%%%%%%%%%%%%%%%%%%%%%%%%%%%%%%%%%%%%%%%%%%%%%%% 

We investigate this mechanism in more detail in Fig. \ref{fig.spin_dynamics} by plotting the full dynamics of the spin-spin correlator [Eq. \eqref{eq.spin_correlator_result}] relative to the spin correlator in the absence of holes. Throughout the entire dynamics, we observe the mentioned flip in correlation for any $d\geq 1$. Furthermore, for weak to intermediate interactions we observe heavy oscillations originating in the density oscillations [Fig. \ref{fig.local_densities}]. At very strong interactions, approaching the free quantum walk of the hole, the relative spin correlation reaches an asymptotic value of $C_S(d,\tau) / C_S^{(0)}(d) \to -1 + 2 / q(q-1)^{d-1}$ at long timescales, $\tau \gg 1 / t$. This is because the hole in this case will always leave any finite region of the Bethe lattice, so that $\sum_{d_h > d} n_h(d_h, \tau) \to 1$ in Eq. \eqref{eq.spin_correlator_result}, while $n_h(0,\tau), n_h(d,\tau) \to 0$. Finally, by carefully analyzing the possible extremal values of $C_S(d,\tau) / C_S^{(0)}(d)$, we find that 
\begin{align}
\frac{C_S(d = 0,\tau)}{C_S^{(0)}(d = 0)} &\in [0, 1],\nn \\
\frac{C_S(d = 1,\tau)}{C_S^{(0)}(d = 1)} &\in [-1 + 1 / q, 0],\nn \\
\frac{C_S(d \geq 2,\tau)}{C_S^{(0)}(d \geq 2)} &\in [-1, 0], 
\label{eq.C_S_limits}
\end{align} 
used as the axis limits on the second axes in Fig. \ref{fig.spin_dynamics}. This result is not limited to the $t$-$J_z$ model investigated in the present paper, but holds in general. It only depends on the depth symmetry of the wave function in Eq. \eqref{eq.exact_many_body_wave_function}, and may be derived by varying $C_S(d,\tau) / C_S^{(0)}(d)$ with respect to the coefficients $C^{(d)}$ of the wave function given that the norm of the wave function is preserved, $\braket{\Psi|\Psi} = 1$. Indeed, we see that $C_S$ does not necessarily explore all the possible values, evident in the cases of $d = 1$ and $d = 2$, Figs. \ref{fig.spin_dynamics}(b) and \ref{fig.spin_dynamics}(c) respectively. 

In this way, we see how we may we characterize both the hole and spin dynamics exactly and very efficiently in these Bethe lattice structures.

\section{Conclusions and outlook}
In this article, we have found exact solutions to the nonequilibrium many-body dynamics and a certain class of eigenstates of a single hole in antiferromagnetic Bethe lattices, described by the fully anisotropic $t$-$J_z$ model. The found eigenstates include the magnetic polaron ground state as well as the ubiquitous string excitations. The latter are in this case exact many-body eigenstates with a vanishing spectral width in contrast to the $t$-$J_z$ model in regular crystal lattices \cite{Wrzosek2021}, as well as in the presence of transverse spin coupling present in the full $t$-$J$ model \cite{Kane1989,Martinez1991}. 

As our methodology yields the full many-body wave function, any correlation function can be calculated very efficiently, illustrated by the investigated spin-spin correlation dynamics. The exact solvability of the model is a result of the fractal self-similarity of Bethe lattices, which we have shown leads to simple recursion relations for the coefficients of the wave function. In particular, the self-similarity of the lattice reduces the number of independent coefficients in the wave function from being exponential to linear in the depth, greatly reducing its complexity. We anticipate that it should be possible to extend the present methodology to nonzero temperatures. In this case, we see two possible routes forward, either by expressing the system dynamics in terms of a full density matrix, or by translating the methodology to finite temperature quantum field theory. In either case, understanding the impact of temperature in these highly idealized lattices may further our understanding of the same phenomena in regular crystal lattices. Finally, the exploration of pairing of two holes is essential to improve our understanding of the mechanisms behind high-temperature superconductivity. The massive simplification found in the Bethe lattices for a single hole in the present paper, may lead to interesting new insights into this scenario, which we hope to explore in the future. 

\acknowledgments
KKN would like to thank Georg M. Bruun for helpful and valuable input on the manuscript. KKN also thanks Jens Havgaard Nyhegn and Thomas Pohl for fruitful discussions. This work has been supported by the Carlsberg Foundation through a Carlsberg Internationalisation Fellowship and the Danish National Research Foundation through the Center of Excellence “CCQ” (Grant agreement no.: DNRF156). 

\appendix

\section{Variational energy at strong interactions} \label{app.strong_eigenstates_d0}
In this section, we derive the variational energy in Eq. \eqref{eq.eigenenergies_order_1}. We use 
\begin{equation}
\psi_n(d) = \sqrt{\lambda} A_n \cdot {\rm Ai}(\lambda(d - d_0) - a_n) = \sqrt{\lambda} \phi_n(x), 
\end{equation}
where $x = \lambda(d - d_0)$, and $\lambda = [(q-2)J_z/(2\sqrt{q-1})t]^{1/3}$. The variational parameter is thus the reference depth $d_0$. Using the equations of motion in Eq. \eqref{eq.equations_of_motion_1}, we obtain the variational energy functional
\begin{align}
\varepsilon_n^{\rm var} =& \sum_{d\geq 1}E_J^{(d)} |\psi_n(d)|^2 - 2\sqrt{q-1} t \sum_{d\geq 1} \psi_n(d)\psi(d+1)   \nn \\
& - 2\sqrt{q}t \cdot \psi_n(0)\psi(1) \nn \\
=& \sum_{d\geq 1}E_J^{(d)} |\psi_n(d)|^2 - 2\sqrt{q-1} t \sum_{d\geq 0} \psi_n(d)\psi(d+1)   \nn \\
& - 2t (\sqrt{q} - \sqrt{q-1}) \cdot \psi_n(0)\psi_n(1).
\label{eq.eps_var_1}
\end{align}
This expression can already be used to numerically determine $d_0$. However, as we shall now show, it is actually possible to determine it analytically. Let us first investigate the contribution from the magnetic energy cost
\begin{align}
\varepsilon_{n,1}^{\rm var} =& \sum_{d\geq 1}E_J^{(d)} |\psi_n(d)|^2 = \frac{J_z}{2} \sum_{d\geq 1} \left[1 + (q-2)d \right] |\psi_n(d)|^2 \nn \\
=&  \frac{J_z}{2} \left[1 + (q-2)d_0 \right] \left[1 - |\psi_n(0)|^2\right] \nn \\
&+ (q-2) \frac{J_z}{2} \sum_{d\geq 1} (d-d_0) |\psi_n(d)|^2.
\label{eq.eps_var_1}
\end{align}
Here, we separate out the contribution from $d_0$. From Eq. \eqref{eq.residue_asymptote}, we get $|\psi_n(0)|^2 \propto J_z / t$. This term, therefore, yields a contribution at order $(J_z / t)^2$ and may be dropped. The term proportional to $(d-d_0)$ will superficially yield a term of order $J_z \cdot l_s^{(n)} \propto (J_z / t)^{2/3}$ [See Eq. \eqref{eq.l_s_n}]. However, as we shall see shortly, there is a corresponding term from the hopping Hamiltonian that cancels this. 

To evaluate this sum, containing $\psi_n(d) \psi_n(d+1)$, we expand $\psi_n(d+1)$ to second order: $\psi_n(d+1) = \psi_n(d) + \partial_d \psi_n + \partial_d^2 \psi_n / 2$. The first of these terms, therefore, contribute with $-2\sqrt{q-1}t \cdot \sum_d |\psi_n(d)|^2 = -2\sqrt{q-1}t$. Further, using the defining differential equation for the Airy function [Eq. \eqref{eq.airy_equation}], we get $\partial_d^2 \psi_n = \lambda^2 [\lambda(d-d_0) - a_n] \psi_n(d)$. Hence, the contribution from the second order derivative is
\begin{align}
\!\!\!\!\!&-\!\sqrt{q-1} t \cdot \lambda^2 \sum_d \psi_n(d) \partial_d^2 \psi_n(d) \nn \\
\!\!\!\!\!&= -\sqrt{q-1} t \cdot \lambda^2 \sum_d [\lambda(d-d_0) - a_n] |\psi_n(d)|^2 \nn \\
\!\!\!\!\!&= \sqrt{q-1}t \cdot a_n \lambda^2 - \frac{(q-2)J_z}{2} \!\sum_{d\geq 0} (d-d_0) |\psi_n(d)|^2. \!\!
\label{eq.eps_var_2_1}
\end{align}
The first term in this expression yields the term at order $(J_z / t)^{2/3}$ to the energy in Eq. \eqref{eq.eigenenergies_order_1}. The second term cancels all contributions in the lower line of Eq. \eqref{eq.eps_var_1} for $d \geq 1$. The remaining term is, thus, proportional to $|\psi_n(d)|^2$, and is therefore proportional to $(J_z/t)^2$ and may be dropped. We now move on to the contribution from the first derivate, $\partial_d \psi_n$. This yields, 
\begin{align}
\sum_d \psi(d)\partial_d\psi_n &= \lambda^2 \sum_{x \geq -\lambda d_0} \phi_n(x) \partial_x \phi_n(x) \nn \\
&= \lambda \sum_{x \geq -\lambda d_0} \Delta x \cdot \phi_n(x) \partial_x \phi_n(x),
\end{align}
using $\Delta x = \lambda$. To evaluate this sum, we transform the sum to an integral using the midpoint rule. This yields
\begin{align}
\!\!&\sum_d \psi(d)\partial_d\psi_n \to \lambda \int_{-\lambda d_0 - \lambda / 2}^\infty {\rm d}x \, \phi_n(x)\partial_x \phi_n(x) \nn \\
\!\!&= \lambda \left[\int_{0}^\infty \!\!\!{\rm d}x \, \phi_n(x)\partial_x \phi_n(x) - \int_{0}^{-\lambda(d_0 + 1/2)} \!\!\!\!\!{\rm d}x \, \phi_n(x)\partial_x \phi_n(x)\right] \nn \\
\!\!&\simeq -\lambda \int_{0}^{-\lambda(d_0 + 1/2)} \!\!\!{\rm d}x \, x \left|[\partial_y \phi_n(y)]_{y = 0}\right|^2 \nn \\
\!\!&= -\frac{\lambda^3}{2}\left(d_0 + \frac{1}{2}\right)^2
\end{align}
The use of the midpoint rule results in the shift of the integration limit from $-\lambda d_0$ to $-\lambda d_0 - \lambda / 2$, i.e. half an interval of $\Delta x$. Consequently, the error made in transforming the sum to an integral is of order $(\lambda(d_0 + 1/2))^3$ and may be neglected. The collected contribution from these hopping terms is, thus, 
\begin{align}
\varepsilon^{\rm var}_{n,2} =&\, -2\sqrt{q-1}t + \sqrt{q-1}t \cdot a_n \cdot \left(\frac{(q-2)J_z}{2\sqrt{q-1}t}\right)^{2/3} \nn \\
&+ \sqrt{q-1}t \cdot \lambda^3\left(d_0 + \frac{1}{2}\right)^2\nn \\
&= \varepsilon_n + (q-2) \frac{J_z}{2} \left(d_0 + \frac{1}{2}\right)^2. 
\label{eq.eps_var_2_final}
\end{align}
Here, we neglect the term $- (q-2)\frac{J_z}{2} \sum_{d\geq 0} (d-d_0) |\psi_n(d)|^2$, cancelling the lower term in Eq. \eqref{eq.eps_var_1}. Finally, we investigate the term proportional to $\psi_n(0)\psi_n(1)$, present due to the fact that the hopping between $d = 0$ and $d = 1$ is fundamentally different than for $d \geq 1$. Expanding the wave functions to lowest order yields 
\begin{align}
\varepsilon^{\rm var}_{n,3} &= - 2t (\sqrt{q} - \sqrt{q-1}) \cdot \psi_n(0)\psi_n(1) \nn \\
&= 2t (\sqrt{q} - \sqrt{q-1}) \cdot \lambda^3 d_0(1-d_0), 
\end{align}
using that $A_n^2 |{\rm Ai}'(-a_n)|^2 = 1$. All in all, we get the variational energy 
\begin{align}
&\varepsilon_n^{\rm var} = \varepsilon_n + \left[1 + (q-2)d_0 \right]\cdot \frac{J_z}{2} \nn \\
&+ (q-2) \left(d_0 + \frac{1}{2}\right)^2 \cdot \frac{J_z}{2}  \nn \\
&+ 2(q-2) \left(\sqrt{\frac{q}{q-1}} - 1\right) d_0(1-d_0) \cdot \frac{J_z}{2},
\label{eq.eps_var_final}
\end{align}
where $\varepsilon_n = -2\sqrt{q-1}t(1 - a_n \lambda^2 / 2)$. The result in Eq. \eqref{eq.eps_var_final} coincides with Eq. \eqref{eq.eigenenergies_order_1}. 

\bibliographystyle{apsrev4-1}
\bibliography{ref_magnetic_polaron}

\end{document}